\documentclass[a4paper,twoside, 12pt]{report}
\usepackage[top=3cm, bottom=3cm, left=3cm, right=3cm]{geometry}
\usepackage[french]{babel}
\usepackage[utf8]{inputenc}
\usepackage[T1]{fontenc}
\usepackage{amsmath, amsfonts, amssymb}
\usepackage{wasysym}
\usepackage{hyperref,url,cite}
\usepackage{graphicx}
\usepackage{subfigure}
\usepackage{xcolor}
\usepackage{geometry}
\usepackage{subfigure}
\usepackage{epsfig}
\usepackage{appendix}
\usepackage{pifont}
\usepackage[nottoc, notlof, notlot]{tocbibind}
\begin{document}
\pagestyle{empty}
\begin{titlepage}
\begin{center}
\includegraphics[width = 3.5cm]{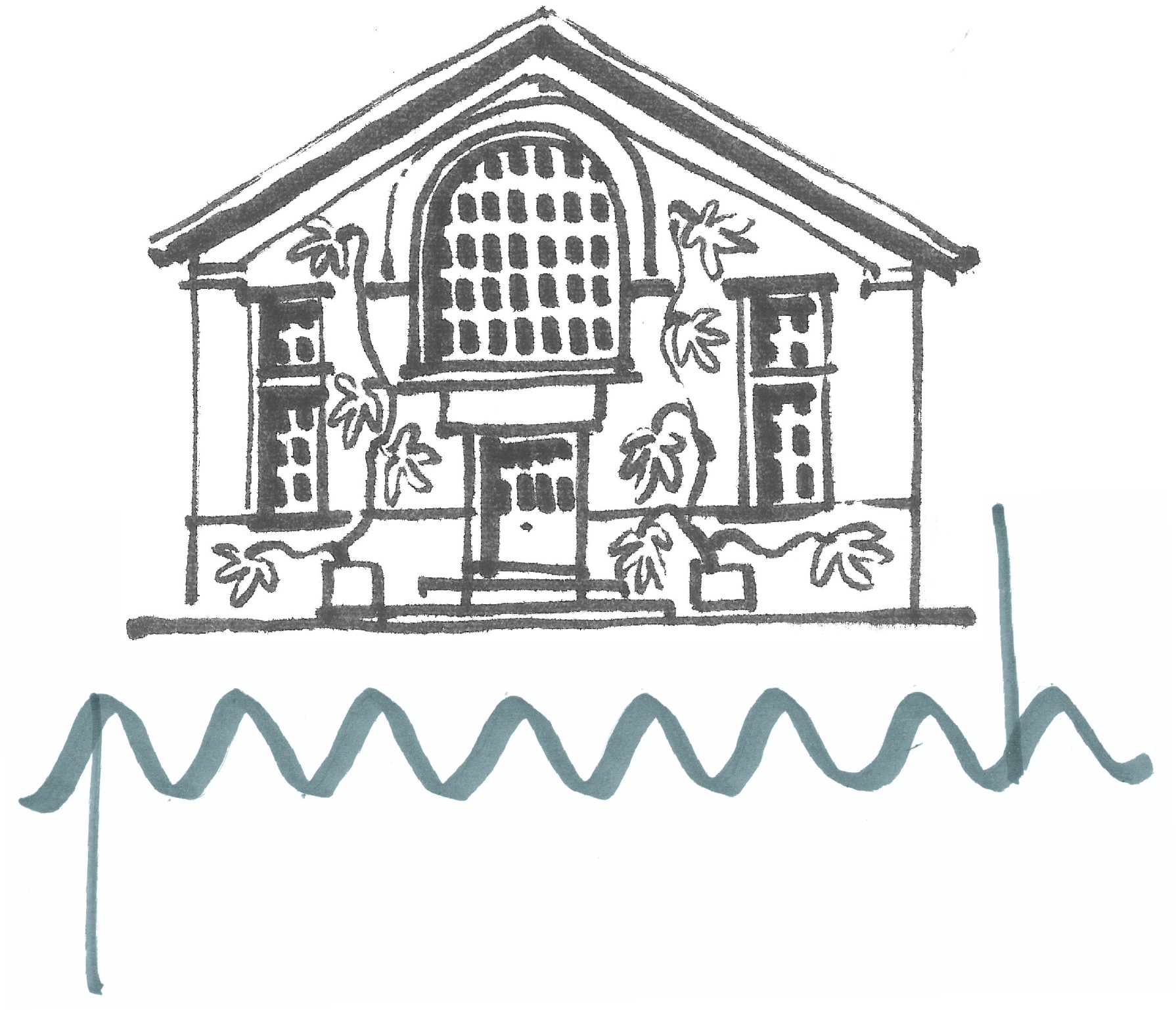} \hspace*{\stretch{1}}\includegraphics[width = 3.5cm]{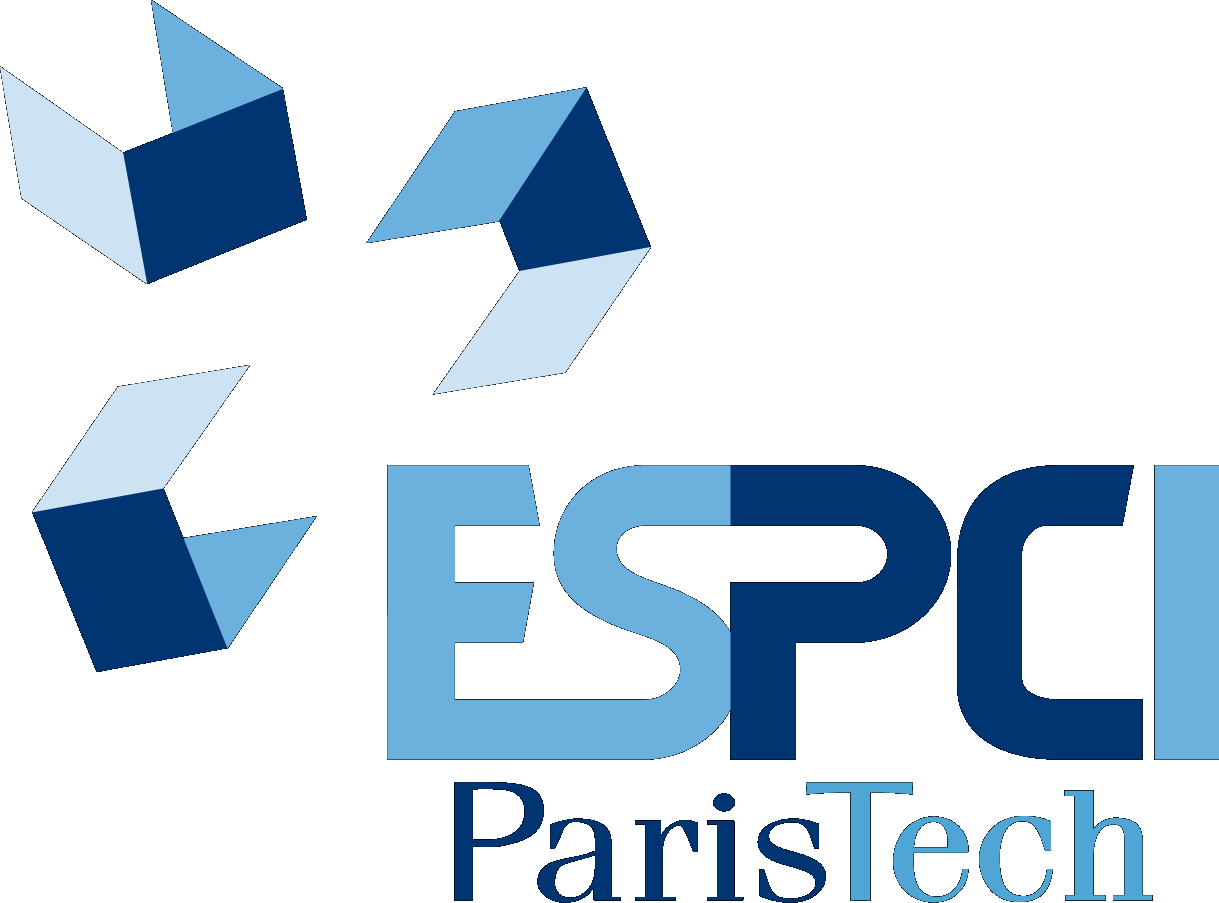} \hspace*{\stretch{1}} \includegraphics[width = 2cm]{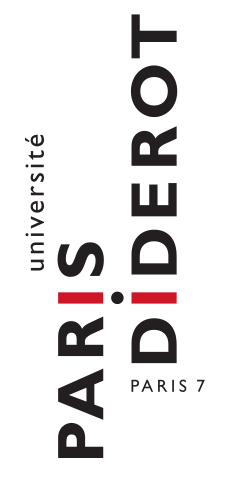}
\end{center}

\vspace*{1cm} 

\begin{center}\bfseries\large
Rapport de stage de première année de Master :
\end{center}

\begin{center}\bfseries\Huge
Étude théorique de l'instabilité de formation des rides éoliennes
\end{center}

\vspace*{0.5cm} 
\hspace*{2cm}\hrulefill\hspace*{2.5cm}
\vspace*{0.5cm} 

\begin{center}\bfseries\Large
Timothé Poulain
\end{center}

\vspace*{0.5cm} 
\hspace*{4.5cm}\hrulefill\hspace*{5cm}
\vspace*{0.5cm} 

\begin{center}\bfseries\large
Laboratoire de Physique et Mécanique des Milieux Hétérogènes

CNRS - ESPCI Paris Tech
\end{center}

\vspace*{\stretch{1}} 

\begin{center}
Directeurs de stage : Bruno Andreotti \& Philippe Claudin

Rapporteur : Julien Derr {\footnotesize(Matériaux et Systèmes Complexes - P7)}
\end{center}

\vspace*{\stretch{1}} 

\begin{center}
\includegraphics[width = 16cm]{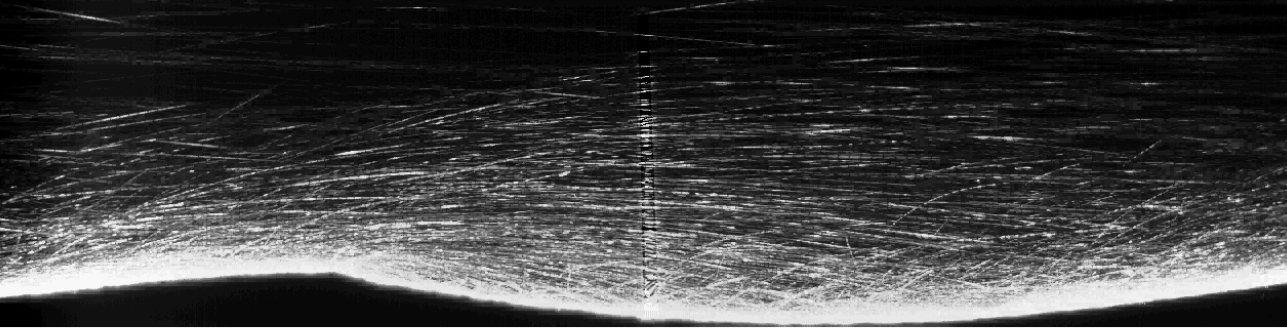}
\end{center}

\vspace*{\stretch{1}} 

\begin{center}
Magistère de Physique Fondamentale \hspace*{\stretch{1}} Année universitaire 2013-2014
\end{center}
\end{titlepage}
\newpage
\null
\newpage
\tableofcontents
\newpage
\null
\newpage
\section*{Remerciements}
\phantomsection
Je remercie Bruno Andreotti pour m'avoir permis de découvrir la recherche théorique à travers l'étude d'une partie des phénomènes qui régissent les milieux granulaires. Je le remercie encore, ainsi que Philippe Claudin, pour leur encadrement tout au long du stage : en plus de m'avoir aidé à éclaircir les problèmes rencontrés, ils ont grandement contribué à développer ma rigueur scientifique. Je remercie également les deux autres stagiaires présents lors de mon séjour, Adeline et Hugo, pour leur soutien et leur aide lorsque mes encadrants officiels n'étaient pas présents.
\newpage
\null
\newpage
\pagestyle{headings}
\chapter*{Introduction}
\begin{figure}
 \centering
 \includegraphics[width=10cm]{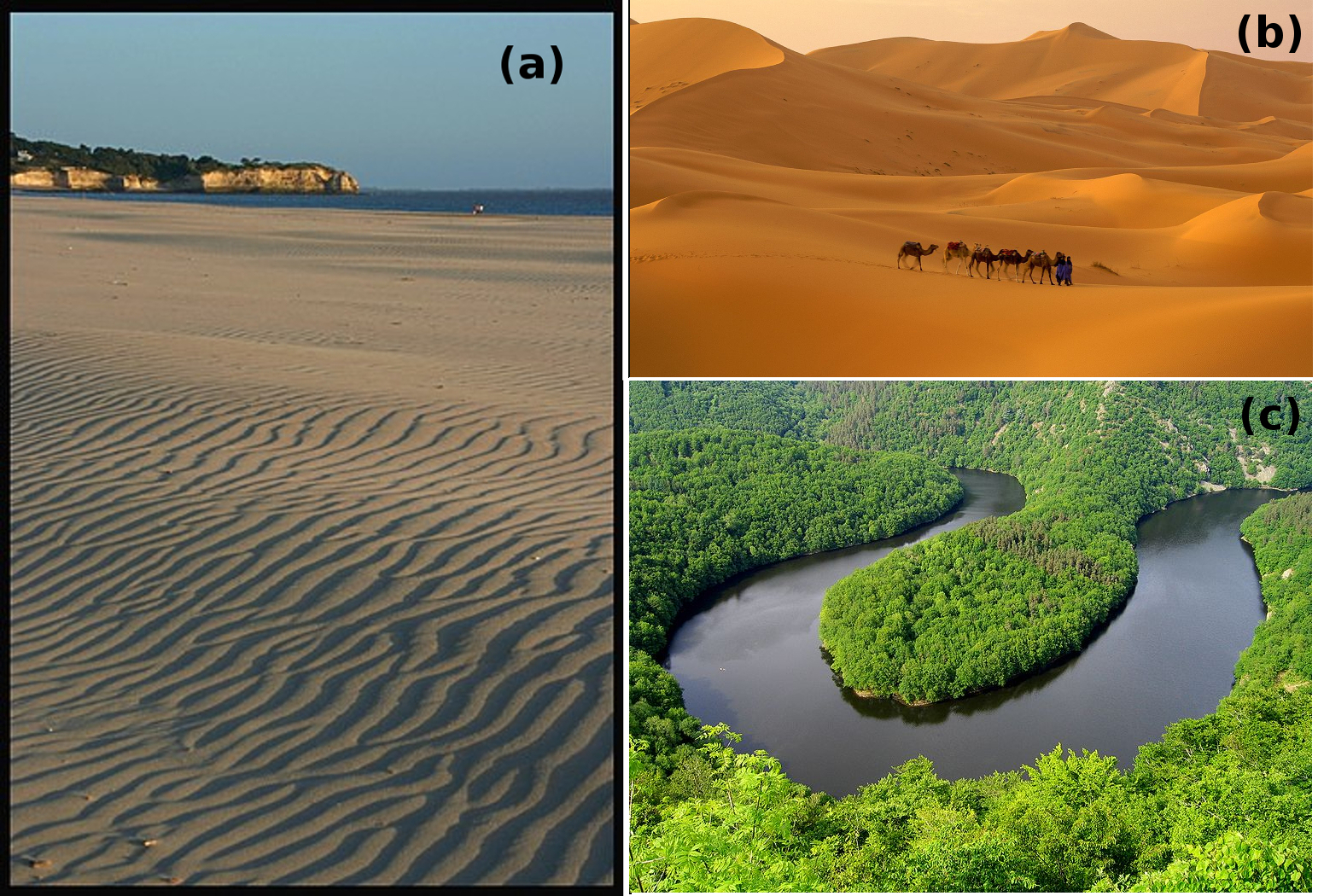}
 \caption{\footnotesize (a) Rides éoliennes à Saint-Georges de Didonne. (b) Dunes de sable dans le désert de Wahiba. (c) Méandre de la Sioule.}
 \label{fig_photos}
\end{figure}

\addcontentsline{toc}{chapter}{Introduction} 
Du transport des grains de sable par le vent au transport fluvial de troncs d'arbre, en passant par le stockage de céréales et les avalanches rocheuses, la grande diversité des systèmes et des processus qui interviennent dans l'étude des milieux granulaires en font une branche de la physique très moderne et fascinante. Fascinante car elle est très polyvalente dans ses thématiques (morphodynamique de la matière en grains, rhéologie des écoulements mixtes fluide-particules, géologie) et parce qu'elle confronte le scientifique à des phénomènes et structures spectaculaires. Moderne car elle offre un large spectre d'applications (transport et conditionnement de denrée alimentaire, manipulation de médicaments, extraction minière) et a pour vocation la compréhension de processus au cœur de problématiques environnementales actuelles (désertification, érosion du littoral). Malgré tout, la grande complexité des phénomènes (turbulence des écoulements, instabilités des structures, nombre important de constituants) rend la compréhension de certains aspects des milieux granulaires très précaire et incomplète. Notamment, que ce soit en géophysique ou dans le cadre d'applications industrielles, il existe de nombreux systèmes, impliquant l’interaction d'un écoulement fluide avec des sédiments, dont la rhéologie est souvent complexe et mal comprise. \medskip
Parmi les thématiques de recherches du laboratoire \textit{Physique et Mécanique des Milieux Hétérogènes}, l'étude des milieux granulaires tient une place importante. Les chercheurs qui y travaillent sont des acteurs actifs du développement de la compréhension des processus sédimentaires. En particulier, Bruno Andreotti et Philippe Claudin, qui ont encadré mon stage, travaillent -- aussi bien en développant des outils numériques qu'en se livrant à l'observation et à l'expérimentation -- sur la formation des rides, des dunes et des méandres (Fig. \ref{fig_photos}).
Souhaitant découvrir la recherche théorique et ayant été intrigué par la physique des milieux granulaires suite à un séminaire organisé par le Magistère de Physique Fondamentale de l'Université Paris Diderot, c'est en toute logique que je me suis tourné vers Bruno Andreotti pour effectuer mon stage.
L'objectif du stage était l'établissement d'un modèle numérique simplifié du transport sédimentaire, en vue d'une étude théorique des instabilités de formation des rides éoliennes.\medskip
Le profil d'un milieu granulaire soumis à un écoulement évolue dans le temps ; on observe que des motifs particuliers s'y dessinent : c'est le cas des dunes et des rides de sable qui adoptent une géométrie (forme, longueur d'onde) caractéristique de la direction et de l'intensité de l'écoulement qui les contraint. Cette évolution de la topographie est associée à des phénomènes d'érosion et de déposition, eux-mêmes liés à la notion de transport sédimentaire. Tous les processus de formation de ces reliefs ne sont pas encore compris : c'est le cas des rides éoliennes. C'est dans ce contexte, et dans la continuité des travaux de Bruno Andreotti et de Philippe Claudin, que se situe mon stage.
Jusqu'alors, les nombreux modèles utilisés pour étudier la formation des rides étaient basés sur l'hypothèse que tous les grains ont la même trajectoire. Soit ces modèles ne sont pas auto-consistants, soit ils conduisent à des solutions instables \cite{bib_transport}. Par ailleurs, les résultats obtenus à partir de ces modèles ne sont pas en accord avec les observations. En particulier, ils prédisent que la longueur d'onde $\lambda$ des rides éoliennes naissantes est indépendante de la vitesse de l'écoulement $u_{*}$, alors que des expériences effectuées en soufflerie montrent que cette longueur d'onde croît avec $u_*$ \cite{bib_experience}.
Orencio Dur\'{a}n, Bruno Andreotti et Philippe Claudin ont récemment effectué des simulations de transport sédimentaire fidèles aux observations \cite{bib_model}. Ces simulations de \textit{dynamique moléculaire} intègrent les trajectoires d'un ensemble de grains transportés par un écoulement cisaillé. Ce modèle tient compte, dans sa description, des lois élémentaires de conservation et de la dynamique, ainsi que des processus les plus fins : collisions interparticulaires, déformation du grain à l'impact, etc. Mais surtout, il prend en compte le caractère probabiliste des trajectoires. Cette différence conceptuelle majeure avec les anciens modèles permet de retrouver les propriétés omises jusqu'à présent. Bien que ce modèle permette de simuler convenablement la formation des rides, sa grande complexité ne permet pas d'isoler clairement la nature des mécanismes d'instabilité qui contrôlent leur formation. On se propose donc de développer un modèle épuré de toute description technique des mécanismes intervenant à l'échelle du grain, dans la limite d'une description cohérente de la nature. Une telle démarche devrait permettre de mieux comprendre le lien étroit qui existe entre l'écoulement et le milieu granulaire, et donc d'en déduire les principaux vecteurs de la formation des rides. 
Finalement, pour souligner l'importance et la légitimité de ces deux approches, on peut associer le modèle de dynamique moléculaire à un outil permettant de réaliser des expériences numériques, et le modèle simplifié à un modèle à proprement parler dans le sens où il permet de déduire les bonnes lois d'échelles et de comprendre plus simplement les processus généraux de la formation des rides éoliennes. \medskip
On commencera par rappeler le formalisme utilisé dans la description du processus d'érosion et du transport sédimentaire avant de décrire le fonctionnement de notre modèle. Enfin, on terminera en exposant les résultats obtenus et en les comparant avec les données connues.
\chapter{Description du transport sédimentaire}
\begin{figure}[h]
 \centering
 \includegraphics[width=10cm]{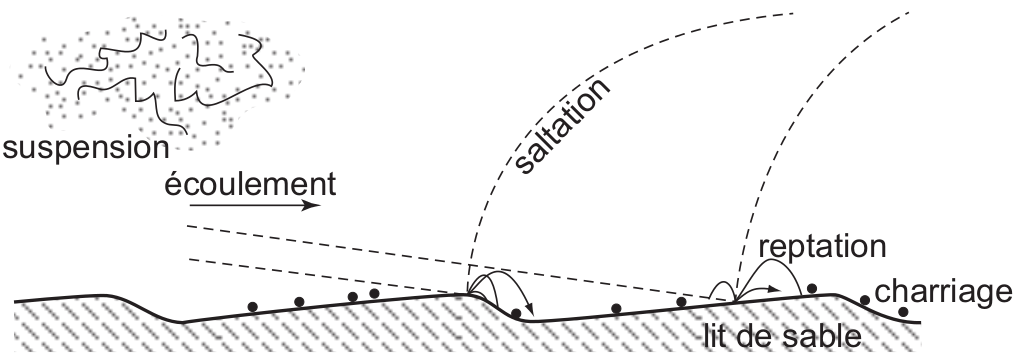}
 \caption{\footnotesize Schéma des différents modes de transport.}
 \label{fig_mode}
\end{figure}

Dans cette partie, on expose les processus qui interviennent dans le transport sédimentaire : une compréhension du lien entre écoulement, lit de sable et flux de matière est nécessaire si on souhaite pouvoir décrire la formation des rides éoliennes. Deux approches de l'étude du transport de particules sont possibles. On peut s'intéresser aux mécanismes à l'échelle du grain, ou bien considérer des flux moyens de matière. C'est la première approche qui a été adoptée dans le modèle de dynamique moléculaire. Pour notre part, on étudiera le transport éolien et tous les processus de formation des rides d'un point de vue statistique - en ne considérant qu'un seul grain, représentatif de tous les autres, effectuant une succession de sauts - ce qui revient à adopter la seconde approche. \medskip
Un grain transporté par un écoulement est soumis à un certain nombre de forces extérieures que l'on peut regrouper, selon leur nature, en trois catégories. On identifie les forces hydrodynamiques (interaction entre les grains et le fluide environnant), les forces de contact (interaction entre particules) et la gravité. Dans le cadre de cette description, on distingue habituellement quatre modes de transport \cite{bib_AFP} (Fig. \ref{fig_mode}). Dans le cas où les forces hydrodynamiques dominent, on parle de \textit{transport en suspension}. Si les particules sont telles que l'attraction gravitationnelle ne peut plus être négligée, on parle de \textit{charge de fond}. Dans ce dernier cas, on observe différents types de trajectoire. Soit la particule effectue une succession de sauts - on parle de \textit{saltation} - soit l'écoulement n'est pas assez puissant pour déloger complètement la particule et celle-ci roule sur le sol - on parle alors de \textit{charriage}. Dans le premier cas, ce sont la gravité et les forces hydrodynamiques qui prévalent, alors que dans le second ce sont la gravité et les forces de contact. Enfin, lorsque l'énergie d'un grain incident est suffisamment grande, celui-ci peut déloger d'autres particules lors de son impact avec le lit. Ce dernier mode de transport est appelé \textit{reptation}.
On peut schématiquement associer ces différents modes de transport - relativement au rapport des masses volumiques des particules et du fluide $\rho_{p} / \rho_{f}$ - à la taille des sédiments qui sont mis en mouvement. Ainsi, les objets les plus fins (cendres ou poussières dans l'air) sont transportés par suspension, alors que les plus grossiers sont transportés par charge de fond : soit par saltation (sable), soit par charriage (galets). \\
Dans notre étude on considère des grains de sable de taille de l'ordre de $100 \mathrm{\mu m}$. Dans ce cas, les modes de transport que l'on est amené à rencontrer sont la saltation et la reptation. C'est ce dernier mode qui est négligé par les anciens modèles.
Il est intéressant de noter que les processus de formation des rides éoliennes sont différents des processus de formation des rides aquatiques. En effet, la viscosité de l'eau étant environ cent fois plus élevée que celle de l'air, un grain de sable transporté par un écoulement sous-marin ou fluvial sera ralenti beaucoup plus rapidement que le même grain dans l'air : le transport par reptation est pratiquement inexistant dans l'eau. On peut s'imaginer que c'est en partie pour cette raison que ce type de trajectoire n'avait pas été considéré jusqu'alors.
\section{Flux et conservation de la matière}
\begin{figure}[h]
 \centering
 \includegraphics[width=10cm]{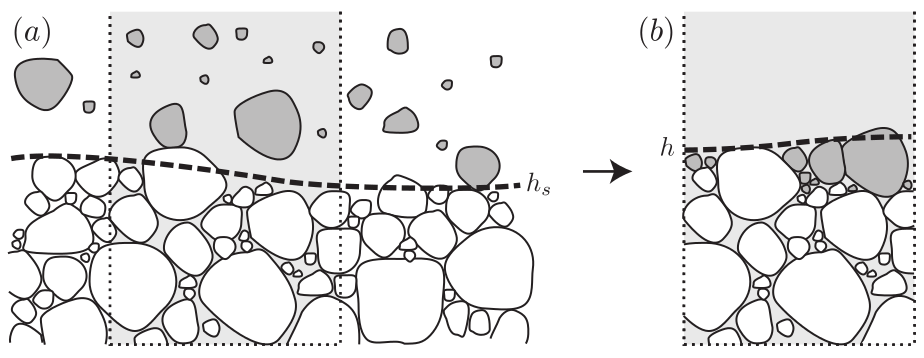}
 \caption{\footnotesize (a) Schéma définissant l'interface $\xi_{s}$ entre le lit statique et la phase composée du fluide et des grains en mouvement. (b) En ramenant virtuellement les grains transportés en surface, de manière à reconstituer un lit effectif de fraction volumique homogène, on définit l'interface $\xi_{d}$.}
 \label{fig_lit}
\end{figure}

Dans le cadre d'une description continue, on peut définir l'érosion et l'accrétion comme des échanges de matière entre deux milieux : le lit sableux et le fluide en écoulement. Il existe deux approches pour définir l'interface $\xi \left( x \right)$ entre ces deux milieux. On peut soit considérer l'interface $\xi_{s} \left(x \right)$ qui sépare les grains statiques des grains mobiles, soit considérer le profil $\xi_{d} \left(x \right)$ qu'aurait le lit s'il n'y avait pas d'écoulement, c'est-à-dire si tous les grains en mouvement étaient déposés sur le sol (Fig. \ref{fig_lit}).
Lorsque les grains sont transportés, la concentration en sédiments en un point du fluide varie et $\partial_{t} \xi_{d}$ correspond à la variation horizontale du nombre de grains à la position $x$.
On note $\vec{q}$ le flux de matière à travers une surface perpendiculaire à l'écoulement (Fig. \ref{fig_flux}.a). La conservation du nombre de grains présents, à un instant donné, dans un volume infinitésimal $V$ centré en $x$ se traduit par la relation intégrale :  
\begin{equation}
 \rho_{p} \Omega_{l} \iiint \partial_{t} \xi_{d} dV = - \oiint q^{i} dS_{i} \, ,
\end{equation}
dont l'équivalent local est :
\begin{equation}
 \rho_{p} \Omega_{l} \partial_{t} \xi_{d} = - \partial_{i} q^{i} \, ,
\end{equation}
où $\Omega_{l}$ est la fraction volumique de grains qui compose le lit.
Pour représenter les quantités de matière échangées verticalement - par érosion et par déposition - entre le fluide et le lit, on introduit les flux d'érosion $\varphi_{\uparrow} \left( z \right)$ et de déposition $\varphi_{\downarrow} \left( z \right)$ comme les masses qui traversent, par unité de temps, une surface unité, horizontale, à l'altitude $z$, respectivement du bas vers le haut et du haut vers le bas (Fig. \ref{fig_flux}.b). Le bilan de matière verticale est calculé par rapport à la variation de $\xi_{s}$ :
\begin{equation}
 \rho_{p} \Omega_{l} \partial_{t} \xi_{s} = \varphi_{\uparrow} - \varphi_{\downarrow} \, .
\end{equation}
Dans notre cas, la couche dynamique - définie par la nappe fluide contenant les grains en mouvement - est confinée à une zone de quelques tailles de grain au-dessus de $\xi_{s}$, beaucoup plus petite que l'expansion spatiale du lit. Ainsi, on peut considérer, en première approximation, $\xi_{s} = \xi_{d}$ \cite{bib_AFP}. 
\section{Saturation du flux de transport}
\begin{figure}[h]
 \centering
 \includegraphics[width=10cm]{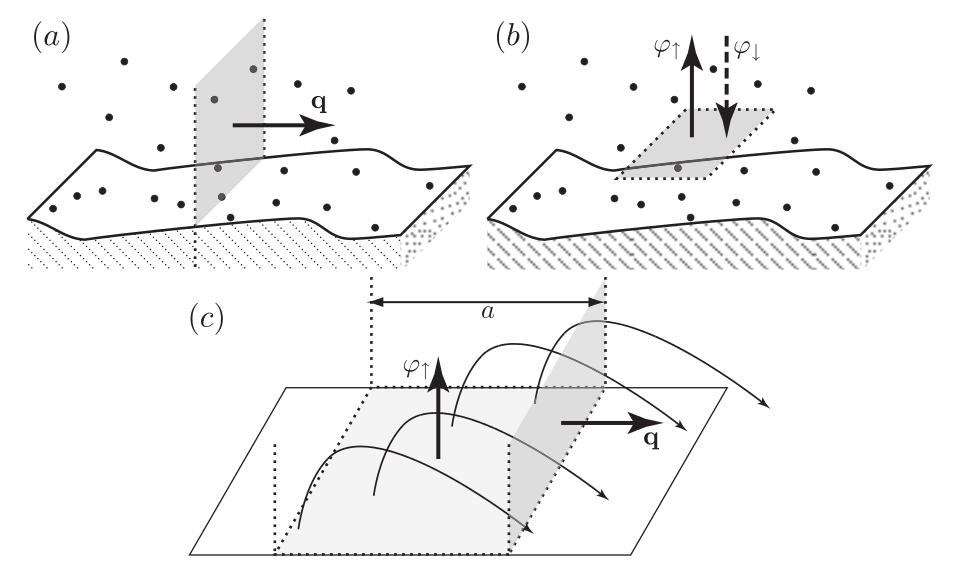}
 \caption{\footnotesize Schémas définissant (a) le flux horizontal $\vec{q}$ et (b) les flux ascendant $\varphi_\uparrow$ et descendant $\varphi_\downarrow$. (c) En régime stationnaire, les flux horizontaux et verticaux sont reliés par la longueur moyenne de saut, $a$.}
 \label{fig_flux}
\end{figure}

Considérons un lit plat soumis à un écoulement permanent. On observe que le nombre de grains transportés par le fluide n'est pas illimité mais est caractérisé par un \textit{flux saturé} noté $q_{sat}$ \cite{bib_rasmussen}. Cette grandeur peut être interprétée comme résultant d'un équilibre entre l'écoulement et le transport de particules et traduit le fait qu'il n'y a globalement ni érosion ni accrétion du lit. Ainsi, le nombre de grains arrachés du lit est, en moyenne, égal au nombre de grains qui s'y déposent : $\varphi_{\uparrow} = \varphi_{\downarrow} = \varphi$. \\
Or, le nombre de grains qui traversent une surface perpendiculaire à l'écoulement après avoir effectué un saut de longueur $\ell$, vérifie la relation :
\begin{equation}
 \label{eq_relation_flux}
 q = \int \ell P \left( \ell \right) \varphi d\ell = a \varphi \, ,
\end{equation}
où $a$ est la longueur moyenne des sauts effectués (Fig. \ref{fig_flux}.c).
En conséquence de (\ref{eq_relation_flux}) et de l'existence d'un flux de transport saturé $q_{sat}$, il existe une valeur limite pour les flux verticaux : $\varphi_{sat}$. On en déduit que ce n'est pas parce qu'un écoulement est très puissant qu'il érodera plus un lit de sédiments qu'un écoulement moins intense. Dans les deux cas - sous la condition que l'écoulement soit suffisamment fort pour pouvoir déloger les grains qui composent le lit - il n'y a globalement ni érosion ni déposition. En revanche, à l'équilibre, plus un écoulement est puissant plus il transporte de particules : $q_{sat}$ est une fonction croissante de la vitesse de cisaillement $u_{*}$ qui s'annule en dessous d'une vitesse de cisaillement seuil ; il en est de même pour $\varphi_{sat}$.
Dans la pratique, lorsque l'écoulement varie dans l'espace ou le temps, le processus de saturation n'est pas instantané. Le flux de sable $\vec{q}$ relaxe exponentiellement vers sa valeur à saturation $q_{sat}$ sur une longueur caractéristique $L_{sat}$ pendant un temps $T_{sat}$. Ne pas tenir compte du régime transitoire peut a priori faire douter de la pertinence de nos résultats. Cependant, de la même manière que l'on peut supposer que l'équilibre stationnaire est atteint après un temps $t > T_{sat}$, on admet que, pour une statistique suffisamment grande de sauts, la contribution du transitoire se confond dans la donnée totale.
\section{Seuil de transport et dynamique}
Une notion importante est celle de \textit{seuil statique de transport}. Un écoulement ne peut éroder un lit sédimentaire que s'il est suffisamment puissant pour compenser la gravité et les forces de cohésion qui s'exercent sur les grains du lit ; en dessous de ce seuil il n'y a ni érosion ni transport. La force hydrodynamique appliquée aux grains est proportionnelle à $\tau^{f} d^{2}$, où $\tau^{f}$ est la contrainte de cisaillement à l'interface grains-fluide. Elle est d'autant plus importante que la pente sur laquelle les grains se situent est raide (relativement au sens de l'écoulement). A contrario si la pente est descendante dans le sens de l'écoulement, celui-ci aura moins de difficulté à déloger les particules. On associe cette contrainte de cisaillement à une vitesse caractéristique, notée $u_{*}$, par la relation $\tau^{f} = \rho_{f} u_{*}^{2}$. On étudie $\tau^{f}$ dans la section \ref{sec_profil}. En terme de vitesse d'écoulement, on peut dire que pour qu'il y ait érosion, il est nécessaire que $u_{*}$ soit supérieure à une vitesse seuil $u_{th}$, où $u_{th}$ est reliée, proche du sol, à la contrainte de cisaillement par $\tau^{f} = \rho_{f} u_{th}^{2}$.
Une fois arraché au lit, un grain (supposé idéalement sphérique de dimension $d$) interagit avec l'écoulement via une force proportionnelle au coefficient de traîné $\mathcal{C}_{x}$ :
\begin{equation}
 \label{eq_frottements}
 \vec{f} = \frac{1}{2} \mathcal{C}_{x} \pi \rho_{f} d^{2} \left( \vec{u} \left( x, z \right) - \vec{v} \right) \mid \vec{u} \left( x, z \right) - \vec{v} \mid \, ,
\end{equation}
où $\vec{u}$ est la vitesse du vent et $\vec{v}$ est la vitesse du grain.
L'équation du mouvement s'écrit alors :
\begin{equation}
 \label{PFD}
 m_{p} \frac{d \vec{v}}{dt} = \frac{1}{2} \mathcal{C}_{x} \pi \rho_{f} d^{2} \left( \vec{u} \left( x, z \right) - \vec{v} \right) \mid \vec{u} \left( x, z \right) - \vec{v} \mid + m_{p} \left( 1 - \frac{\rho_{f}}{\rho_{p}} \right) \vec{g} \, .
\end{equation}
Lorsque le grain est délogé, il est accéléré par l'écoulement et, si l'énergie emmagasinée pendant son vol est suffisamment importante, il peut, lors de l'impact avec le lit, rebondir en éjectant d'autres particules. Ainsi, de plus en plus de grains sont érodés - à la force de cisaillement s'ajoute l'inertie des grains transportés - et le flux de particules incidentes augmente. Cependant, ce processus d'amplification ne persiste pas indéfiniment mais sature en raison de la rétroaction du transport sur l'écoulement. La vitesse de l'écoulement dans la couche de transport diminue jusqu'à ce que chaque grain n'expulse, en moyenne, plus qu'un seul grain \cite{bib_transport}. Le fait que chaque grain transporté peut contribuer à l'érosion du lit implique que la vitesse $u_{th}$ décroît avec le nombre de grains transportés et se stabilise à l'équilibre. Cet autre seuil de transport est appelé \textit{seuil dynamique}. \medskip
C'est l'ensemble du processus dans lequel les grains rebondissent, sont accélérés par le vent et éjectent d'autres grains lors de la collision avec le lit, qui est appelé la saltation ; les grains associés sont appelés les \textit{saltons}. C'est le mode principal du transport sédimentaire par le vent. Le second mode de transport, la reptation, est quant à elle associée aux grains qui sont éjectés par les saltons et qui s'arrêtent presque directement après leur premier saut. On appelle \textit{reptons} les particules associées à ce mode de transport. Comme on le verra plus tard, ce sont les reptons qui sont majoritairement responsables de la formation des rides éoliennes.
\section{Profil de l'écoulement}
\label{sec_profil}
Les mesures effectuées en soufflerie indiquent que l'évolution de la longueur d'onde des rides éoliennes sur le sable dépend linéairement de la vitesse du vent \cite{bib_experience}. Pour rendre notre modèle réaliste, il est nécessaire que l'on décrive le plus correctement possible le profil de vitesse de l'écoulement proche du lit. On justifie dans cette partie, à l'aide de considérations hydrodynamiques, le choix effectué pour décrire le profil de vitesse du vent.
Soit un écoulement stationnaire incompressible de vitesse $\vec{u}\left(x,z \right)$. On considère, dans un premier temps, un lit plat. Les équations qui décrivent l'écoulement sont les équations de Navier-Stokes (\ref{eq_NS}) et l'équation de continuité (\ref{eq_continuite}) :
\begin{equation}
 \label{eq_NS}
 \rho_{f} \left[ \partial_{t} u_{i} + u_{j} \partial^{j} u_{i} \right] = - \partial_{i} p + \mu \left[ \partial_{j} \partial^{j} u_{i} + \frac{1}{3} \partial_{i} \partial^{j} u_{j} \right]
\end{equation}
\begin{equation}
 \label{eq_continuite}
 \partial^{i} u_{i} = 0
\end{equation}
Dans l'air, à une altitude $z = 1 \mathrm{m}$ et pour une vitesse du vent $\mid \vec{u} \mid = 1 \mathrm{m}.\mathrm{s}^{-1}$, le nombre de Reynolds qui caractérise l'écoulement est de l'ordre de $10^{6} \gg 1$. Autrement dit, le régime des écoulements que l'on sera amené à considérer est turbulent.
Les grandeurs qui caractérisent un tel écoulement sont soumises à des variations aléatoires. Ainsi, on peut décomposer la vitesse $\vec{u}\left(x,z \right)$ et la pression $p\left(x,z \right)$ en une composante moyenne $\overline{*}$ et une composante de fluctuation $*^{'}$. La composante de fluctuation traduit l'intensité de la turbulence et est définie telle que sa valeur moyenne soit nulle.
\begin{equation}
 \label{eq_decomposition}
 \left\lbrace
  \begin{array}{lll}
   u_{i} & =& \bar{u}_{i} + u_{i}^{'} \\
   p & = & \bar{p} + p^{'}
  \end{array}\right.
\end{equation}
En effectuant la moyenne statistique de l'équation obtenue en injectant (\ref{eq_decomposition}) dans (\ref{eq_NS}), on obtient les équations de Navier-Stokes pour la composante moyenne de la vitesse de l'écoulement. On procède de la même manière pour l'équation de continuité et on obtient finalement le nouveau jeu d'équations :
\begin{eqnarray}
  \rho_{f} \left[ \partial_{t} \bar{u}_{i} + \bar{u}_{j} \partial^{j} \bar{u}_{i} + \overline{u_{j}^{'} \partial^{j} u_{i}^{'}} \right] & = & - \partial_{i} \bar{p} + \mu \partial_{j} \partial^{j} \bar{u}_{i} \\
  \partial^{i} \bar{u}_{i} & = & 0
\end{eqnarray}
que l'on peut encore écrire :
\begin{eqnarray}
 \label{eq_NSdecomposition}
 \rho_{f} \left[ \partial_{t} \bar{u}_{i} + \partial^{j} \left( \bar{u}_{j} \bar{u}_{i} \right) \right] & = & - \partial_{i} \bar{p} + \partial^{j} \left[ \mu \partial_{j} \bar{u}_{i} - \rho_{f} \overline{u_{j}^{'} u_{i}^{'}} \right] \\
 \partial^{i} \bar{u}_{i} & = & 0
\end{eqnarray}
Lorsque l'on considère les équations de Navier-Stokes pour la composante moyenne du champs de vitesse, il s'ajoute aux forces visqueuses une pseudo-force qui se traduit par une contrainte de cisaillement turbulente $\tau^{f}_{ij} = \rho_{f} \overline{u_{i}^{'} u_{j}^{'}}$, appelée tenseur de Reynolds. Ce tenseur de contraintes turbulentes est a priori non nul et est représentatif des corrélations qui existent entre les fluctuations des différentes composantes de la vitesse de l'écoulement.
La connaissance de $\tau^{f}_{ij}$ permet d'évaluer la contrainte de cisaillement induite par l'écoulement sur le milieu granulaire et d'en déduire le profil de vitesse du vent. Pour arriver à ce résultat, on considère, comme Prandtl, l'écoulement composé de \textit{balles fluides}. Elles correspondent aux groupements fluides qui se forment en conséquence des mouvements tourbillonnaires. Ces balles fluides, qui se propagent sur une échelle caractéristique appelée \textit{longueur de mélange}, contribuent aux échanges de quantité de mouvement entre les différentes couches fluides. Elles permettent ainsi de caractériser la contrainte exercée entre deux couches voisines de l'écoulement.
On considère un écoulement turbulent plan $\vec{u} = u_{x} \left( z \right) \vec{e}_{x}$. D'après (\ref{eq_NSdecomposition}), la contrainte tangentielle au voisinage de la paroi est :
\begin{equation}
 \label{eq_contrainte}
 \tau = \mu \frac{\partial \overline{u}_{x}}{\partial z} - \rho_{f} \overline{u_{x}^{'} u_{z}^{'}} \, .
\end{equation}
À un petit déplacement vertical $l$ doit correspondre une petite variation $\Delta u_{x}$ de $u_{x}$ ; ce qui se traduit mathématiquement par la relation $\Delta u_{x} = l \left( \partial u_{x} / \partial z \right)$. Le postulat de Prandtl a été d'admettre que si $l$ est du même ordre de grandeur que la longueur de mélange, alors $\Delta u_{x} \sim \sqrt{\langle u_{x}^{'2} \rangle}$. En admettant de plus que $\sqrt{\langle u_{x}^{'2} \rangle} \sim \sqrt{\langle u_{z}^{'2} \rangle}$, on en déduit : 
\begin{equation}
 \sqrt{\langle u_{x}^{'2} \rangle} \sqrt{\langle u_{z}^{'2} \rangle} \sim l^{2} \frac{\partial \overline{u}_{x}}{\partial z} \, .
\end{equation}
D'où, en considérant le coefficient de corrélation $C_{u_{x}^{'} u_{z}^{'}} = \frac{\overline{u_{x}^{'} u_{z}^{'}}}{\sqrt{\langle u_{z}^{'2} \rangle \langle u_{z}^{'2} \rangle}}$ qui vaut approximativement $-0.4$ \cite{bib_hydro}:
\begin{equation}
 \tau = \rho_{f} \left(\nu + \xi \right) \frac{\partial \overline{u}_{x}}{\partial z} \, ,
\end{equation}
où $\xi = - C_{u_{x}^{'} u_{z}^{'}} l^{2} \frac{\partial \overline{u}_{x}}{\partial z}$ est la viscosité cinématique turbulente. En régime turbulent, $\mid \xi \mid \gg \nu$, d'où :
\begin{equation}
 \tau \simeq \rho_{f} l^{2} \left( \frac{\partial \overline{u}_{x}}{\partial z} \right)^{2} \, .
\end{equation}
On suppose de plus, qu'au voisinage de la paroi, la longueur de mélange est proportionnelle à la distance z : $l = \kappa z$ - avec $\kappa$ la constante phénoménologique de Von Kármán - et que le frottement ne peut s'écarter notablement du frottement à la paroi.
Intégrer cette équation permet finalement de trouver l'équation qui régit le profil de vitesse de l'écoulement :
\begin{equation}
 \label{eq_profil_plat}
 u_{x}(z) = \frac{u_{*}}{\kappa} \ln \left(\frac{z}{z_{0}} \right) \, ,
\end{equation}
où $z_{0}$ est une constante d'intégration homogène à une longueur, appelée \textit{rugosité hydrodynamique}, et $u_{*}$ est une vitesse caractéristique de cisaillement définie par $\tau^{f}_{xz} = \rho_{f} u_{*}^{2}$.
La rugosité hydrodynamique $z_{0}$ est, par définition, la hauteur à laquelle la vitesse semble s'annuler, si on prolonge au sol le profil logarithmique. Dans le cas d'un sol lisse, $z_{0}$ correspond à l'épaisseur de la sous-couche limite visqueuse. Dans le cas d'un sol rugueux, si la sous-couche limite visqueuse est plus petite que la rugosité du lit, alors $z_{0}$ est déterminée par cette dernière. Pour un lit plat de grains statiques, on trouve expérimentalement $z_{0} \sim d/30$ \cite{bib_rugosite}.
Pour un lit sinusoïdal de faible amplitude, la formule (\ref{eq_profil_plat}) s'adapte, en première approximation, en déjaugeant la hauteur du lit au point $x$, $Z(x)$, à l'altitude $z$ :
\begin{equation}
 \label{eq_profil_sinusoidal}
 u_{x}(z) = \frac{u_{*}}{\kappa} \ln \left(\frac{z - Z(x)}{z_{0}} \right) \, .
\end{equation}
\chapter{Description du modèle numérique} 
\begin{figure}[h]
 \centering
 \includegraphics[width = 8cm]{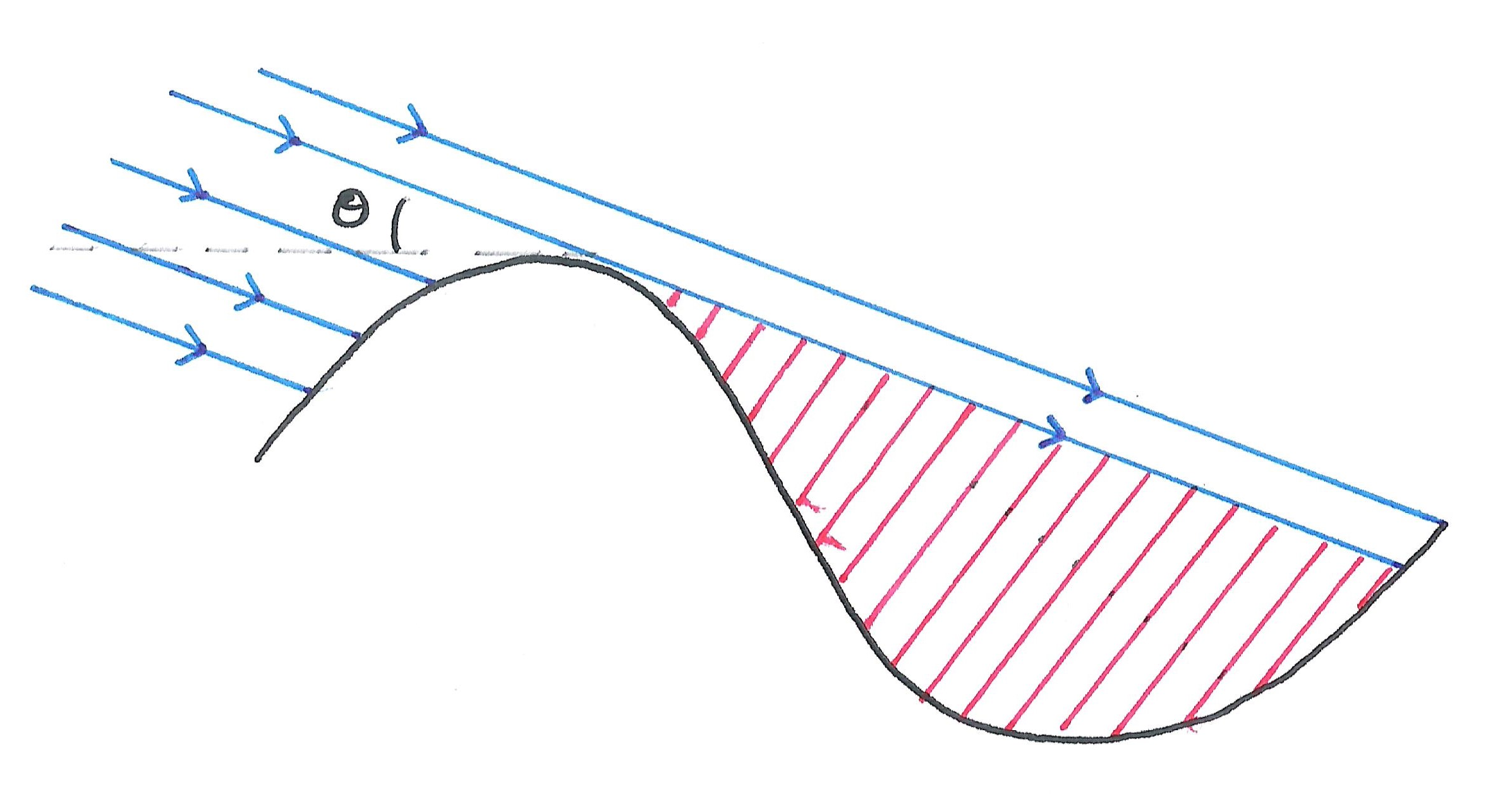}
 \caption{\footnotesize Le taux de déposition des grains sur un lit ondulé soumis à un écoulement cisaillé n'est pas constant mais varie suivant la phase du lit. Ainsi, la région hachurée est-elle moins exposée au flux incident que celles contenues dans l'angle solide des grains.}
 \label{fig_argument_geometrique}
\end{figure}

On expose dans cette partie le fonctionnement de notre modèle du transport sédimentaire, dans lequel les trajectoires successives d'un unique grain sont calculées. \medskip
On suppose que le temps caractéristique d'évolution du profil du lit est beaucoup plus long que le temps de vol du grain, de sorte que le relief n'est pas modifié pendant le calcul d'une trajectoire du grain.
On considère le cas du lit plat ($Z \left( x \right) = Z_{0}$) comme l'état de référence et on note $\varphi_{0}$ le taux de déposition de grains sur ce lit. Comme le lit est en tout point identique et coïncide à tout instant avec le profil initial, $\varphi_{0}$ doit être constant pour un vent donné - il tombe globalement le même nombre de grains en chaque point du lit. De même, on note $\psi_{0} \left( \ell \right)$ le taux de déposition conditionné par la longueur de saut $\ell$. Cette grandeur, dont l'intégrale sur $\ell$ est par définition égale à $\varphi_{0}$, correspond à la distribution des longueurs de saut $\mathcal{P} \left( \ell \right)$. Cette distribution - que l'on normalise à 1 - peut être déterminée expérimentalement à partir de relevés effectués en soufflerie \cite{bib_model}.
On suppose que chaque trajectoire du grain est initiée par une vitesse dont le module est tiré aléatoirement dans une distribution $P \left( v \right)$ en accord avec $\mathcal{P} \left( \ell \right)$ (cf. section \ref{sec_distribution}). Ce choix est motivé par la nature statistique de notre description du transport sédimentaire : on ne considère qu'un seul grain représentatif de tous les autres. Par ailleurs, on impose - en première approximation - l'angle entre le vecteur vitesse et l'horizontale à $\frac{\pi}{4}$. Dans cette configuration, un grain qui décolle de la position $x_{\uparrow}$ atterri, après avoir effectué un saut de longueur $\ell_{0}$, en $x_{\downarrow} = x_{\uparrow} + \ell_{0}$. \medskip
Soit un lit sinusoïdal $Z \left( x \right) = Z_{0} + Z_{1} \left( x \right)$ proche du lit plat, c'est-à-dire dont le rapport d'aspect - défini comme le rapport de l'amplitude sur la longueur d'onde - est petit devant 1. On pose $Z_{1} \left( x \right) = \zeta \cos \left( kx \right)$, la petite variation du profil du lit autour de $Z_{0}$. Afin d'alléger la notation, on fixe l'altitude de référence $Z_{0} = 0$.
L'étude des mécanismes du transport sur un relief ondulé est intéressante car on peut associer ce dernier à un lit sur lequel des rides éoliennes se sont déjà formées. Ainsi, le calcul de la répartition des grains sur une telle étendue permet de connaître où les grains se déposent le plus pour une longueur d'onde donnée.
Dans le cas d'un lit faiblement ondulé, les trajectoires sont modifiées et, au premier ordre en la perturbation $k \zeta$, la variation de la longueur de saut est donnée par :
\begin{equation}
 \label{eq_variation_saut}
 \ell - \ell_{0} = \zeta \cos \left( k x_{\downarrow} \right) \mathcal{S} \, , 
\end{equation}
où $\mathcal{S}$ est fonction de $k \ell$. \\
On comprend intuitivement que cette modification de la longueur de saut provient de la géométrie du lit : à vitesse initiale identique, la longueur de saut d'un grain n'aura pas la même valeur que sa trajectoire aboutisse sur une crête ou dans un creux. Cependant, l'argument géométrique n'est pas la seule explication à la modification de la distribution $\mathcal{P} \left( \ell \right)$ : il faut également tenir compte de la modulation du profil de l'écoulement induite par l'ondulation du lit. Ces deux contributions sont discutées en \ref{sec_theorique}.
Pour des raisons analogues, les réponses des flux verticaux $\varphi_{1} \left( x \right)$ et $\psi_{1} \left( x, \ell \right)$ à une petite variation du profil du lit sont sinusoïdales. On s'en convainc facilement pour $\varphi_{1}$ à partir de considérations géométriques. Comme le montre le schéma (Fig. \ref{fig_argument_geometrique}), pour un flux incident faisant un angle $\theta$ avec l'horizontale, il existe une région - dont l'aire est fonction de $\theta$ - qui ne reçoit, statistiquement, presque aucune particule. Les zones sur lesquelles les grains se déposent majoritairement sont celles qui sont les plus exposées à l'écoulement. Cette remarque implique également qu'il existe un déphasage $\phi$ entre $\varphi_{1} \left( x \right)$ et $Z \left( x \right)$ ; l'observation de l'avancement spatial des rides au cours du temps confirme cette idée que les grains ne tombent pas principalement au sommet des crêtes. En effet, comme les grains se déposent davantage juste avant la crête que juste après - relativement au sens de l'écoulement - les rides avancent par le sommet en prenant l'allure de vagues.

Au premier ordre en $k \zeta$, le taux de déposition s'écrit :
\begin{equation}
 \label{eq_varphi}
 \varphi \left( x \right) = \varphi_{0} + \varphi_{1} \left( x \right) \, ,
\end{equation}
avec :
\begin{equation}
 \label{eq_varphi1}
 \varphi_{1} \left( x \right) = \varphi_{0} k \zeta A^{\varphi} \cos \left( kx + \phi^{\varphi} \right) \, .
\end{equation}
On a de même :
\begin{equation}
 \label{eq_psi}
 \psi \left( x, \ell \right) = \psi_{0} \left( \ell \right) + \psi_{1} \left( x, \ell \right) \, ,
\end{equation}
avec :
\begin{equation}
 \label{eq_psi1}
 \psi_{1} \left( x, \ell \right) = \psi_{0} \left( \ell \right) k \zeta A^{\psi} \left( \ell \right) \cos \left( kx + \phi^{\psi} \left( \ell \right) \right) \, .
\end{equation}
La distribution $\psi \left( x_{\downarrow}, \ell \right)$, qui est définie comme le nombre de grains qui tombent par unité de temps entre $x_{\downarrow}$ et $x_{\downarrow} + dx_{\downarrow}$ après avoir effectué un saut dont la longueur est comprise entre $\ell$ et $\ell +d\ell$, satisfait l'équation de conservation :
\begin{equation}
 \label{eq_conservation}
 \psi \left( x_{\downarrow}, \ell \right) d\ell dx_{\downarrow} = \mathcal{P} \left( \ell_{0} \right) d\ell_{0} \varphi_{\uparrow} \left( x_{\uparrow} \right) dx_{\uparrow} \, ,
\end{equation}
où $\varphi_{\uparrow}$ est le taux d'érosion que l'on a supposé égal au taux de déposition $\varphi_{\downarrow} = \varphi$ (régime permanent). D'après (\ref{eq_varphi}) et (\ref{eq_psi}), l'équation (\ref{eq_conservation}) s'écrit, à l'ordre zéro :
\begin{equation}
 \psi_{0} \left( x_{\downarrow}, \ell \right) = \mathcal{P} \left( \ell \right) \varphi_{0} \left( x_{\uparrow} \right) \, .
\end{equation}
Le profil du lit étant $2 \pi$-périodique, on peut considérer, d'un point de vue statistique, qu'il revient au même qu'un grain atterrisse en $x_{\downarrow}$ ou en $x_{\downarrow} + 2 \pi n$, $n \in \mathbb{N}$. On sera donc amené à étudier $\varphi$ et $\psi$ fonctions de la phase du lit $\chi \equiv kx [2 \pi]$ plutôt que fonctions de $x$. \medskip
On expose, dans une première section, la méthode d'intégration numérique utilisée pour calculer les trajectoires du grain. On établit ensuite la distribution des vitesses initiales de saut $P \left( v \right)$ utilisée dans le modèle. Dans une troisième section, on détaille les calculs numériques des flux verticaux. Puis, on aborde le calcul des relations de dispersion qui caractérisent le lit. Enfin, on termine en établissant les relations théoriques de $\mathcal{A}^{\psi}$ et $\phi^{\psi}$ qui nous serviront à valider notre modèle dans le chapitre suivant. \medskip
Il est à noter que l'on a choisi de décrire la perturbation avec $k \zeta$ plutôt qu'avec $\zeta / \lambda $. Ce choix de convention nous affranchit des reports successifs, dans les calculs, d'un facteur $2 \pi$.
\section{Algorithme d'intégration des trajectoires}
\subsection{Adimensionnement des équations du mouvement}
Une étude granulométrique de l'étendue sableuse que l'on considère permet de connaître la distribution en taille des grains qui la composent. En première approximation, nous pouvons considérer un ensemble de grains de sable sphériques tous de dimension $d$ : la taille la mieux représentée de l'échantillon. \\ 
Soit une particule se déplaçant à la vitesse $\vec{v}$ dans un écoulement d'air dont le profil de vitesse est donné par $\vec{u}(x,z)$. Elle est soumise, lors de son déplacement, au champ gravitationnel et à des forces de frottement, et vérifie l'équation (\ref{PFD}). Afin de pouvoir résoudre numériquement l'équation du mouvement, il est nécessaire de l'adimensionner. Pour ce faire on introduit les variables sans dimensions $\vec{v}^{*} = \left( \vec{v} / U \right)$, $\vec{x}^{*} = \left( \vec{x} / L \right)$ et $\vec{t}^{*} = \left( \vec{t} / T \right)$, où U, L et T sont des grandeurs caractéristiques du problème. Dans le cas de notre étude, il est commode de parler en terme de grain. C'est pourquoi on choisit $L = d$, $U=\sqrt{gd}$ et $T = \sqrt{\left( d / g \right)}$. Le problème nécessite alors deux paramètres de contrôle seulement : la vitesse de l'écoulement divisée par $\sqrt{gd}$ et le rapport des masses volumiques $\rho_{p}/\rho_{f}$. En divisant toute l'équation par le pré-facteur du membre de gauche obtenu on trouve :
\begin{equation}
 \label{eq_PDF_adimension}
 \frac{d \vec{v}^{*}}{dt^{*}} = 3 C_{x} \frac{\rho_{f}}{\rho_{p}} \left(\frac{\vec{u}}{\sqrt{gd}}-\vec{v}^{*}\right) \left|\frac{\vec{u}}{\sqrt{gd}}-\vec{v}^{*}\right|-\left(1- \frac{\rho_{f}}{\rho_{p}}\right) \vec{e}_{z} \, .
\end{equation}
\subsection{Implémentation au modèle}
\begin{figure}[h]
 \centering
 \includegraphics[width = 10cm]{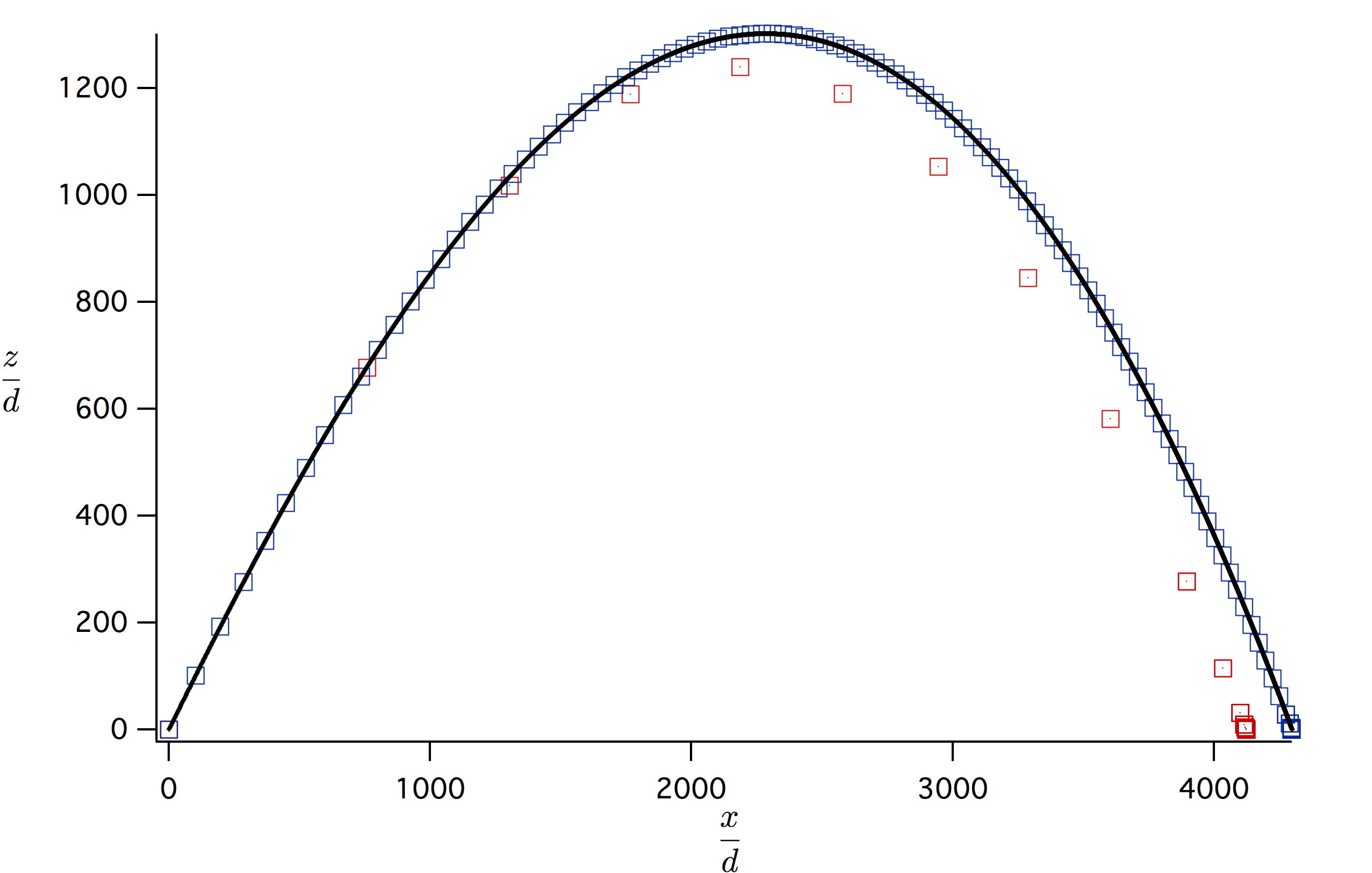}
 \caption{\footnotesize Détermination du pas d'intégration à l'aide de trois trajectoires : en rouge $h = 10$, en bleu $h = 1$ et en noir $h = 0.01$.}
 \label{fig_RK4}
\end{figure}
\begin{figure}[h]
 \centering
 \includegraphics[width = 15cm, height = 3.5cm]{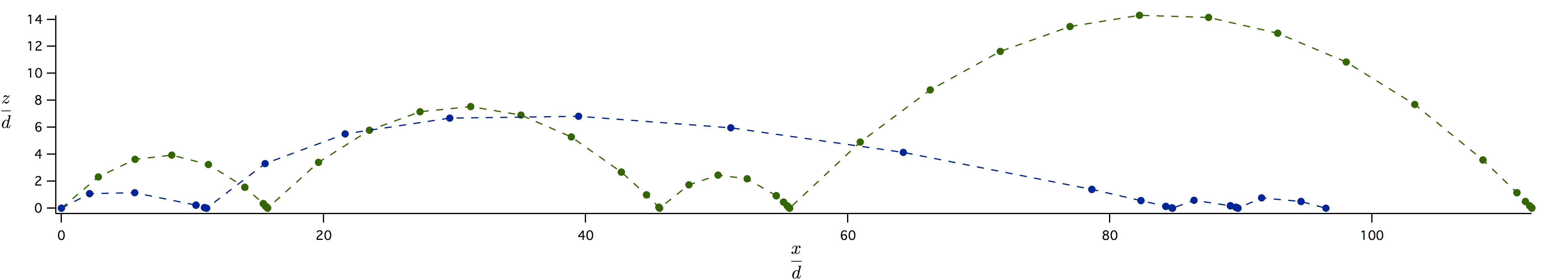}
 \caption{\footnotesize Trajectoires d'un grain effectuant une succession de sauts pour deux vents différents : en vert $u_* / \sqrt{gd} = 0$ et en bleu $u_* / \sqrt{gd} \simeq 3.2$}
 \label{fig_trajectoire}
\end{figure}

On intègre cette équation différentielle du second ordre par la méthode Runge Kutta 4. Cette méthode de calculs permet de résoudre les équations du type $\left( dy / dx \right) = f \left( x, y \right)$ à l'aide de quatre estimations. La première estimation est celle d'Euler\footnote{$y_{i+1} = y_{i} + h f \left( x_{i}, y_{i} \right)$ et $x_{i+1} = x_{i} + h$, avec $h$ le pas d'intégration} pour un demi-pas $(h/2)$. En repartant du point $\left( x_{i}, y_{i} \right)$, on applique de nouveau la méthode d'Euler, pour un demi-pas, en utilisant la valeur de la dérivée calculée à l'étape précédente. On obtient une nouvelle valeur de $y_{i+1}$. La troisième estimation est calculée de la même manière que la seconde mais à partir de la deuxième estimation de la dérivée. Toujours à partir du point de départ et en utilisant la valeur de la dérivée trouvée à la troisième étape, on calcule la quatrième estimation de $y_{i+1}$ en utilisant la méthode d'Euler pour un pas $h$. Finalement, une moyenne pondérée de ces quatre estimations permet de calculer une valeur de $y_{i+1}$. Ces étapes se résument par l'équation :
\begin{equation}
 y_{i+1} = y_{i} + \frac{h}{6}\left( f(x_i,y_i) + 2f(x_i+\frac{h}{2},y_i+\frac{k_1}{2}) + 2f(x_i+\frac{h}{2},y_i+\frac{k_2}{2}) + f(x_i+h,y_i+k_3) \right) \, ,
 \nonumber
\end{equation}
où $k_1 = h f \left( x_i, y_i \right)$, $k_2 = h f \left( x_i + \frac{h}{2}, y_i + \frac{k_1}{2} \right)$, $k_3 = h f \left( x_i + \frac{h}{2}, y_i + \frac{k_2}{2} \right)$ et $k_4 = h f \left( x_i + h, y_i + k_3 \right)$.
On ramène l'équation (\ref{eq_PDF_adimension}) à un système d'équations différentielles du premier ordre en posant :
\begin{equation}
 \left\lbrace
 \begin{array}{lll}
  \dot{x}^{*} & = & v_{x}^{*}\\
  \enskip \\
  \dot{z}^{*} & = & v_{z}^{*}\\
  \enskip \\
  \dot{v}_{x}^{*} & = & 3 C_{x} \frac{\rho_{f}}{\rho_{p}} \left(\frac{u_{x}}{\sqrt{gd}}-v^{*}_{x}\right) \sqrt{\left(\frac{u_{x}}{\sqrt{gd}}-v^{*}_{x}\right)^{2}+\left(\frac{u_{z}}{\sqrt{gd}}-v^{*}_{z}\right)^{2}}\\
  \enskip \\
  \dot{v}_{z}^{*} & = & 3 C_{x} \frac{\rho_{f}}{\rho_{p}} \left(\frac{u_{z}}{\sqrt{gd}}-v^{*}_{z}\right) \sqrt{\left(\frac{u_{x}}{\sqrt{gd}}-v^{*}_{x}\right)^{2}+\left(\frac{u_{z}}{\sqrt{gd}}-v^{*}_{z}\right)^{2}}-\left(1- \frac{\rho_{f}}{\rho_{p}}\right)
 \end{array}\right.
\end{equation}
Pour obtenir les résultats les plus précis possible tout en optimisant les temps de calcul, il faut utiliser un pas d'intégration ni trop grand ni trop petit. En calculant la trajectoire d'un grain, avec les mêmes conditions initiales, pour différentes valeurs du pas d'intégration, on a déterminé que $h = 1$ convient (Fig. \ref{fig_RK4}). La figure (Fig. \ref{fig_trajectoire}) illustre deux exemples de trajectoire calculée pour deux vents différents : $u_* / \sqrt{gd} = 0$ et $u_* / \sqrt{gd} \simeq 3.2$. Comme on pouvait s'y attendre, les grains soumis à un écoulement sont portés par le vent et leurs trajectoires sont allongées dans le sens du vent.
\section{Distribution des vitesses initiales de saut}
\label{sec_distribution}
\begin{figure}[h]
 \centering
 \includegraphics[width = 12cm]{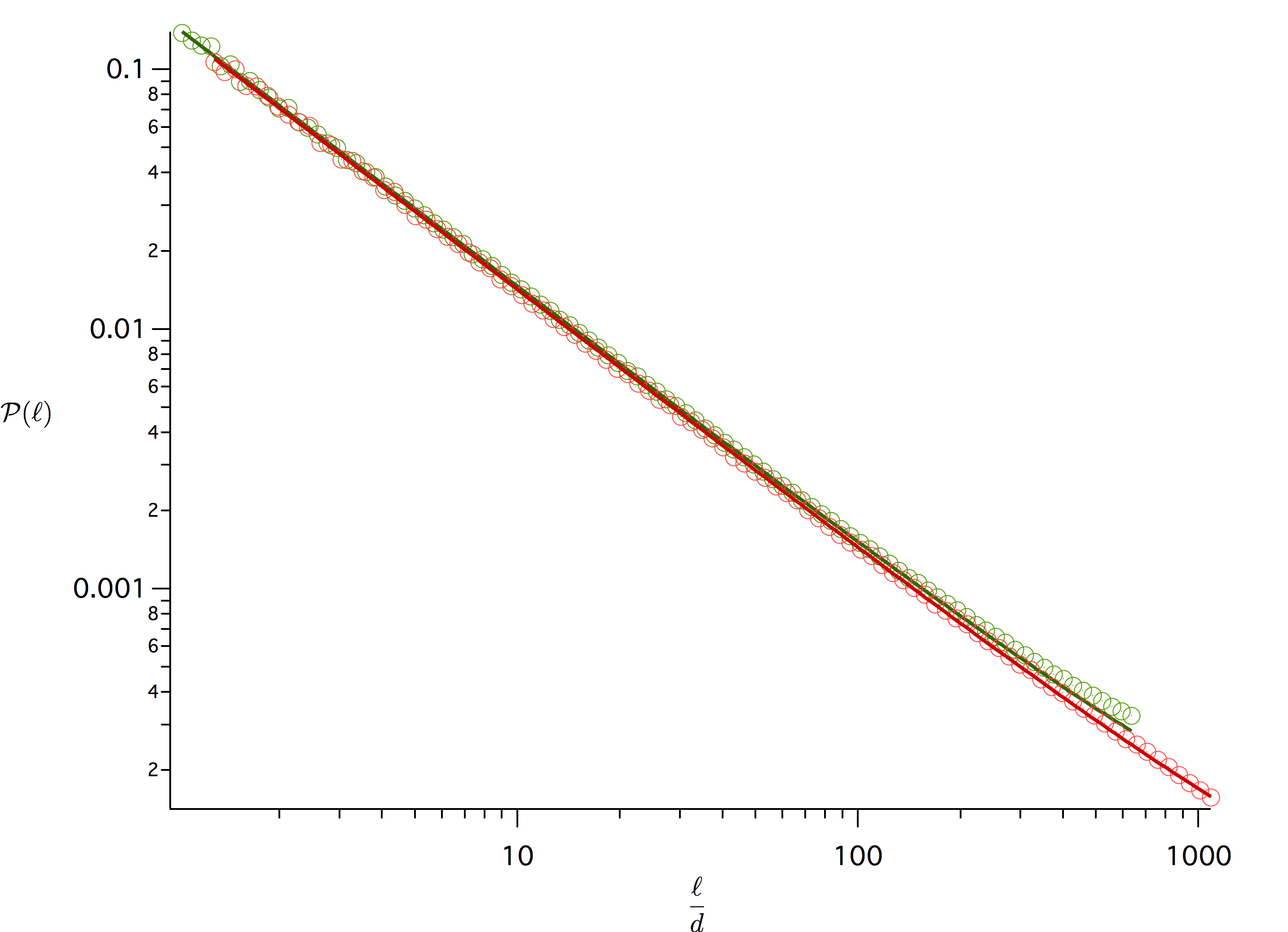}
 \caption{\footnotesize Distribution des longueurs de saut avec (ronds rouges) et sans (ronds verts) vent. Les traits pleins correspondent aux ajustement par une loi de puissance inverse.}
 \label{fig_distsaut}
\end{figure}

Les observations indiquent que, sur lit plat, les longueurs de saut effectué par des grains de sable, soumis à un écoulement cisaillé, sont distribuées suivant une loi de puissance $\mathcal{P} \left( \ell_{0} \right) \propto 1 / \ell_{0}$ et sont à valeur dans l'intervalle $\left[ \ell_{min} ; \ell_{max} \right]$ . Pour que cette distribution soit normalisée à 1 - ie $\int_{\ell_{min}}^{\ell_{max}} \mathcal{P} \left( \ell_{0} \right) d\ell_{0} = 1$ - il faut introduire une constante de normalisation $\alpha = \left[\ln\left( \ell_{max} \right) - \ln \left( \ell_{min} \right) \right]^{-1}$.
Pour être en accord avec les observations, le modèle doit permettre de retrouver statistiquement cette distribution. L'unique paramètre de contrôle sur lequel on peut jouer pour retrouver ce résultat est la vitesse initiale du grain. \medskip
Dans le cas où le terrain n'est soumis à aucune contrainte de cisaillement (vent nul), un grain de sable, éjecté du lit avec une vitesse initiale $\vec{v}$, décrit une trajectoire parabolique dont la longueur de saut $\ell \sim \frac{\mid \vec{v} \mid}{g}^{2}$. Pour que la distribution des vitesses initiales $P \left( v \right)$ soit en accord avec la distribution $\mathcal{P} \left( \ell_{0} \right)$, il faut qu'elle vérifie l'égalité $P \left( v \right) dv = \mathcal{P} \left( \ell_{0} \right) d\ell_{0}$. Dans l'approximation des trajectoires paraboliques, on trouve $P \left( v \right) = 2 \alpha / v$.
Numériquement, nous avons accès à une distribution $p \left( y \right)$ uniforme entre 0 et 1. On peut, comme précédemment, établir la relation entre $v$ et $y$ à partir de l'équation $P \left( v \right) dv = p \left( y \right) dy$. On trouve alors, en intégrant et en imposant les conditions aux limites $v \left( 0 \right) = v_{min}$ et $v \left( 1 \right) = v_{max}$, l'équation qui donne la vitesse du grain au point d'émission : 
\begin{equation}
 \label{eq_dist_vitesse}
 v \left( y \right) = v_{min} e^{\left[ \ln \left( v_{max} \right) - \ln \left( v_{min} \right) \right] y} \, .
\end{equation}
On vérifie cette expression en calculant, pour un grand nombre de trajectoires, les longueurs de saut $\ell_{0}$ d'un unique grain. On utilise pour le calcul $\ell_{min} = 1 d$ et $\ell_{max} = 1000 d$ ; ces valeurs impliquent une constante de normalisation $\alpha \approx 0.145$. La figure (Fig. \ref{fig_distsaut}) représente la distribution $\mathcal{P} \left( \ell_{0} \right)$ calculée avec et sans vent. Pour homogénéiser la répartition des grains entre les bins, on a utilisé des boîtes de taille variant exponentiellement avec $\ell_{0}$.
On remarque que la distribution avec vent peut être ajustée par une fonction inverse dont le pré-facteur est égal à $\alpha$ à $0.25 \%$ près ; ce qui valide la distribution des vitesses initiales de saut choisie (\ref{eq_dist_vitesse}). On remarque également qu'elle explore de plus grandes valeurs de $\ell$ que la distribution sans vent, ce qui provient du fait qu'un grain soumis à un écoulement turbulent est accéléré par celui-ci.
\section{Modulation du taux de déposition}
On détaille, dans un premier temps, le calcul du taux de déposition. Il est utile de calculer cette quantité parce que l'on peut facilement la mesurer expérimentalement. Il suffit d'utiliser un seau, que l'on place dans le sable, puis compter la quantité de grains qui y tombent. On obtient ainsi le flux vertical de sable tombé, par unité de surface et de temps, en un point du lit. De plus, ces quantités nous permettront de vérifier la consistance de nos calculs en comparant la courbe du taux de déposition calculée directement à partir des données brutes à celle tracée à partir du calcul des paramètres $A^{m}$ et $\phi^{m}$ ($m \in \left\lbrace \varphi ; \psi \right\rbrace$). Ces quatre derniers paramètres - et plus particulièrement $A^{\psi}$ et $\phi^{\psi}$ - sont les quantités vraiment importantes à calculer puisqu'elles permettent d'établir les relations de dispersion qui définissent l'évolution du lit. On appellera, respectivement, module et phase de la perturbation les $A^{m}$ et $\phi^{m}$. 
\subsection{Calcul numérique de $\varphi$ et $\psi$}
Le taux de déposition, ramené à une longueur d'onde du lit, $\varphi \left( \chi \right)$, correspond au nombre de grains qui tombent en une certaine phase de la première ondulation. Pour calculer numériquement cette quantité à partir des caractéristiques de la trajectoire (longueur de saut, phase du profil à l'impact), il suffit de diviser l'intervalle $\left[0;2\pi\right[$ en $N_{\chi}$ segments de longueurs $\Delta \chi = \left(2\pi / N_\chi \right)$ égales et de compter le nombre de fois que le grain tombe entre $\chi$ et $\chi + \Delta \chi$. Le calcul de $\psi \left( \chi, \ell \right)$ est similaire. Il suffit d'ajouter la contrainte supplémentaire qu'une valeur de $\psi$ correspond au nombre de fois que le grain est tombé dans l'intervalle $\left[ \chi ; \chi + \Delta \chi \right]$ après avoir effectué un saut de longueur comprise entre $\ell$ et $\ell + \Delta \ell$ ; avec $\Delta \ell = \left( \ell_{max} - \ell_{min} \right) / N_\ell$.
\subsection{Calcul du module et de la phase de la perturbation}
En développant (\ref{eq_varphi}) dans (\ref{eq_varphi1}), on fait apparaître la décomposition de Fourier de $\varphi$ :
\begin{equation}
 \label{eq_fourier_varphi}
 \varphi \left(\chi \right) = \varphi_{0} + \varphi_{0} A^{\varphi} k \zeta \left[ \cos \left(\phi^{\varphi} \right) \cos \left(\chi \right)- \sin \left(\phi^{\varphi} \right) \sin \left(\chi \right) \right] \, ,
\end{equation}
qui se limite au premier mode car on s'est restreint à l'ordre 1 du développement en perturbation.
Les projections de $\varphi$ sur $\cos \left(\chi \right)$ et $\sin \left(\chi \right)$, que l'on note respectivement $C_{\varphi}$ et $S_{\varphi}$, permettent d'établir deux nouvelles relations qui définissent $A^{\varphi}$ et $\phi^{\varphi}$ (\ref{eq_parametres_varphi}). Ce sont ces deux quantités (\ref{eq_projection_varphi}) qui seront calculées lors des simulations.
\begin{equation}
 \label{eq_projection_varphi}
 \left\lbrace
 \begin{array}{lll}
  C_{\varphi} & = & \frac{1}{2 \pi} \int_{0}^{2\pi} \varphi \left(\chi \right) \cos \left(\chi \right) d\chi \\
  \qquad \\
  S_{\varphi} & = & \frac{1}{2 \pi} \int_{0}^{2\pi} \varphi \left(\chi \right) \sin \left(\chi \right) d\chi
 \end{array}\right.
\end{equation}
$\int_{0}^{2\pi} \varphi_{0} \cos \left(\chi \right) d\chi$ étant nulle, le calcul de $C_{\varphi}$ et de $S_{\varphi}$ se limite à celui des intégrales de la composante sinusoïdale du taux de déposition. On peut écrire, compte tenu du second terme de (\ref{eq_fourier_varphi}) :
\begin{equation}
 \label{eq_parametres_varphi}
 \left\lbrace
 \begin{array}{lll}
  A^{\varphi} & = & \frac{2}{\varphi_{0} k \zeta} \sqrt{C_{\varphi}^{2} + S_{\varphi}^{2}} \\
  \qquad \\
  \phi^{\varphi} & = & \arctan(\frac{-S_{\varphi}}{C_{\varphi}})
 \end{array}\right.
\end{equation}
On obtient des relations analogues pour $\psi$ :
\begin{equation}
 \label{eq_parametres_psi}
 \left\lbrace
 \begin{array}{lll}
  A^{\psi} (\ell) & = & \frac{2}{\psi_{0}(\ell) k \zeta} \sqrt{C_{\psi}^{2}(\ell) + S_{\psi}^{2}(\ell)} \\
  \qquad \\
  \phi^{\psi} (\ell) & = & \arctan(\frac{-S_{\psi}(\ell)}{C_{\psi}(\ell)})
 \end{array}\right.
\end{equation}
\section{Calcul des relations de dispersion}
\label{sec_dispersion}
\subsection{Description préliminaire}
Pour la première partie de notre étude, on a considéré un unique grain capable d'effectuer un grand nombre de sauts, et on en a déduit - statistiquement - des valeurs pour $A^{\psi}$ et $\phi^{\psi}$. Lors de chaque impact, le grain était réémis avec une vitesse aléatoire $\vec{v}$ et le relief n'était pas modifié au cours du temps. Si on souhaite étudier les relations de dispersion qui caractérisent le lit, on doit tenir compte de l'évolution de son profil : un grain peut, selon son énergie, soit se déposer, soit rebondir en éjectant d'autres particules. On se persuade facilement qu'un grain très énergétique a plus de chance d'éjecter des particules lors de son impact avec le lit qu'un grain de plus faible énergie. Or plus un grain est accéléré par l'écoulement, plus il emmagasine de l'énergie et plus sa vitesse est grande. Des études ont montré \cite{bib_ejection} que la vitesse d'éjection est, en moyenne, proportionnelle à la vitesse d'impact : ceci signifie qu'une partie de la quantité de mouvement transportée par le grain incident est transférée au{\footnotesize(x)} grain{\footnotesize(s)} éjecté{\footnotesize(s)}. C'est pourquoi, la vitesse d'impact étant proportionnelle à la longueur de saut, on associe cette dépendance énergétique à $\ell$.
Pour faciliter les calculs, on adopte la notation complexes : 
\begin{equation}
 A^{\psi} \left( \ell \right) \zeta \cos \left( \chi + \phi^{\psi} \right) \rightarrow \mathcal{A}^{\psi} \left( \ell \right) Z \left( \chi \right) \, ,
\end{equation}
avec :
\begin{equation}
 \left\lbrace
 \begin{array}{llll}
  Z \left( \chi \right) & = & \zeta e^{i \chi} \\
 \mathcal{A}^{\psi} \left( \ell \right) & = & A^{\psi} \left( \ell \right) e^{i \phi^{\psi} \left( \ell \right)}
 \end{array}\right.
\end{equation}
\medskip
L'évolution du lit est caractérisée par un \textit{taux de croissance} $\sigma (k)$ et une \textit{pulsation} $\omega (k)$. Lorsqu'il y a déposition, $Z \left( x \right)$ augmente et $\sigma \left( k \right) > 0$ ; lorsqu'il y a érosion, $\sigma \left( k \right) < 0$. On a alors $Z \left( x, t \right) = \zeta e^{i \chi - i \omega t + \sigma t}$. Physiquement, $\sigma$ et $\omega$ sont proportionnels  à la différence des nombres de grains qui arrivent et qui repartent. Un grain qui arrive en $x_{\downarrow} = x_{\uparrow} + \ell$ après son $n^\text{ème}$ saut est un grain qui était arrivé en $x_{\uparrow}$ lors des son $(n-1)^\text{ème}$ saut. Et comme, en régime permanent, le taux de déposition est globalement égal au taux d'érosion, le nombre de grains qui sont arrivés en $x_{\uparrow}$ au $(n-1)^\text{ème}$ saut peuvent être considérés comme ceux dont la $n^\text{ème}$ trajectoire est partie de $x_{\uparrow}$. Ainsi, $\psi \left( x_\uparrow, \ell \right) = \psi \left( x_\downarrow, \ell \right) e^{i k \ell}$ et on peut écrire :
\begin{equation}
 \frac{\partial Z}{\partial t} = \int_{0}^{+\infty} \psi \left( x_\downarrow, \ell \right) d\ell - \int_{0}^{+\infty} \psi \left( x_{\uparrow}, \ell \right) d\ell \label{description2} \, ,
\end{equation}
d'où on déduit, en notant $\Re[]$ et $\Im[]$ les parties réelle et imaginaire :
\begin{eqnarray}
 \sigma \left( k \right) & = & \Re \left[ k \varphi_{0} \int_{0}^{+\infty} \mathcal{P} \left( \ell \right) \mathcal{A}^{\psi} \left( \ell \right) \left( 1 - e^{i k \ell} \right) d\ell \right] \label{eq_sigma2} \, , \\
 \omega \left( k \right) & = & \Im \left[ - k \varphi_{0} \int_{0}^{+\infty} \mathcal{P} \left(\ell \right) \mathcal{A}^{\psi} \left(\ell \right) \left( 1 - e^{i k \ell} \right) d\ell \right] \label{eq_omega2} \, .
\end{eqnarray}
\subsection{Modélisation de l'évolution du lit}
Dans l'optique d'établir un modèle simplifié du transport sédimentaire, on introduit la quantité $\left( 1 - \varepsilon \left( \ell \right) \right)$, appelée \textit{capacité de remplacement}, qui correspond à la quantité moyenne de grains réémis après chaque impact. D'après cette définition, la fonction $\varepsilon$ - dont on se sert pour représenter l'action de l'écoulement mixte fluide-particules sur le relief - doit vérifier les deux conditions suivantes : $\varepsilon > 0$ si il y a déposition et $\varepsilon \leq 0$ si il y a érosion ; avec $\left( 1 - \varepsilon \left( \ell \right) \right) > 0$. Dans la pratique, on calcul les projections $C_{\psi, \varepsilon}$ et $S_{\psi, \varepsilon}$ de la même manière que les grandeurs définies à l'équation (\ref{eq_parametres_psi}), à la différence que l'on attribue, à chaque rebond, un poids $w_{n} = \prod\limits_{i = 1}^{n} \left( 1 - \varepsilon \left( \ell_{i} \right) \right) w_{0}$ au grain. $w_{0}$ représente le nombre initial de grains incident, qui, dans notre cas, est égal à 1. Autrement dit, on ne compte plus simplement le nombre de fois que le grain tombe dans l'intervalle $\left[ \chi ; \chi + \Delta \chi \right]$ après avoir effectué un saut de longueur comprise entre $\ell$ et $\ell + \Delta \ell$, mais on calcule une somme pondérée par la fraction de grains réémis - c'est-à-dire par l'énergie du grain.
\begin{figure}[h]
 \centering
 \includegraphics[width = 5cm]{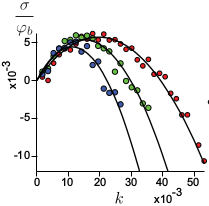}
 \caption{\footnotesize Taux de croissance calculé numériquement par le modèle de dynamique moléculaire pour différents vents. En rouge : $u = 3 u_{th}$, en vert : $u = 4 u_{th}$, en bleu : $u = 5 u_{th}$. Le trait noir représente un ajustement par une fonction $ak - bk^2$}
 \label{fig_sigma_moleculaire}
\end{figure}
Toute la difficulté réside dans la détermination d'une fonction $\varepsilon$ qui permette de retrouver le bon comportement pour $\sigma$ et $\omega$. C'est-à-dire, dans le cas de $\sigma$, une partie déstabilisante linéaire pour les petites valeurs de $k$ puis une partie stabilisante proportionnelle à $-k^{2}$ pour les grands vecteurs d'onde (Fig. \ref{fig_sigma_moleculaire}). Un critère important pour contraindre $\varepsilon$ est la condition qu'il y ait sur lit plat, en moyenne (pour un nombre $N$ de sauts suffisamment grands), autant de grains éjectés que déposés :
\begin{equation}
 \frac{1}{N} \sum \limits_{n} w_n = w_0 \, ,
\end{equation}
que l'on peut réécrire en terme de $\varepsilon$ comme :
\begin{equation}
 \label{lit_plat}
 \int_{0}^{+ \infty} \psi_{0} \left( \ell \right) \varepsilon \left( \ell \right) d\ell = 0 \, .
\end{equation}
La conservation de la matière induit une relation entre l'évolution temporelle de $Z \left(x, t \right)$ et $\varepsilon$ :
\begin{eqnarray}
 \frac{\partial Z}{\partial t} & = & \int_{0}^{+\infty} \psi \left( x,\ell \right) \varepsilon \left( \ell \right) d\ell \label{eq_evolution} \\
 \Rightarrow (\sigma - i\omega) & = & k \int_{0}^{+\infty} \psi_{0} \mathcal{A}^{\psi} \left( \ell \right) \varepsilon \left(\ell \right) d\ell \label{description1}
\end{eqnarray}
d'où on déduit, par identification des parties réelle et imaginaire, les expressions du taux de croissance et de la pulsation du lit :
\begin{eqnarray}
 \sigma \left( k \right) & = & \Re \left[ k \varphi_{0} \int_{0}^{+\infty} \mathcal{P} \left(\ell \right) \mathcal{A}^{\psi} \left( \ell \right) \varepsilon \left(\ell \right) d\ell \right] \label{eq_sigma1} \, , \\
 \omega \left( k \right) & = & \Im \left[ - k \varphi_{0} \int_{0}^{+\infty} \mathcal{P} \left(\ell \right) \mathcal{A}^{\psi} \left( \ell \right) \varepsilon \left(\ell \right) d\ell \right] \label{eq_omega1} \, .
\end{eqnarray}
Pour que les deux descriptions (\ref{eq_sigma2}) et (\ref{eq_sigma1}) du calcul de $\sigma$ soient équivalentes, il faut - en développant $\mathcal{A}^\psi$ au première ordre en $\varepsilon$ : $\mathcal{A}^\psi = \mathcal{A}_{0}^{\psi} + \varepsilon \mathcal{A}_{1}^{\psi}$ - que les deux composantes de $\mathcal{A}^\psi$ vérifient :
\begin{equation}
 \mathcal{A}_1^{\psi} = \frac{\mathcal{A}_0^{\psi}}{1 - e^{i k \ell}} \, .
\end{equation}
On commencera notre étude avec une fonction créneau conditionnée par les propriétés établies dans les paragraphes précédents :
\begin{equation}
 \label{eq_creneau}
 \left\lbrace
 \begin{array}{lllll}
  \varepsilon \left( \ell \right) & = & \varepsilon_{d} & , & \text{si $\ell > \ell_{c}$} \\
  \varepsilon \left( \ell \right) & = & \varepsilon_{g} & , & \text{si $\ell < \ell_{c}$} \\
  1 - \varepsilon \left( \ell \right) & \geq & 0
 \end{array}\right.
\end{equation}
où $\ell_c$, appelée \textit{longueur de coupure}, correspond au point de bascule entre le régime de saltation et de reptation. C'est encore la longueur de saut représentative de l'énergie minimale nécessaire pour qu'un grain incident éjecte d'autres particules lorsqu'il entre en collision avec le lit.
\section{Détermination théorique de $A^{\psi}$ et $\phi^{\psi}$}
\label{sec_theorique}
On établit dans cette section une expression analytique de $\mathcal{A}^{\psi}$ qui servira de référence à nos résultats. Pour cela, on étudie les deux contributions à la modulation des longueurs de saut dont on a déjà discuté lors de l'établissement de l'équation (\ref{eq_variation_saut}).
\subsection{Effet géométrique}
\begin{figure}[h]
 \centering
 \includegraphics[width = 6.5cm]{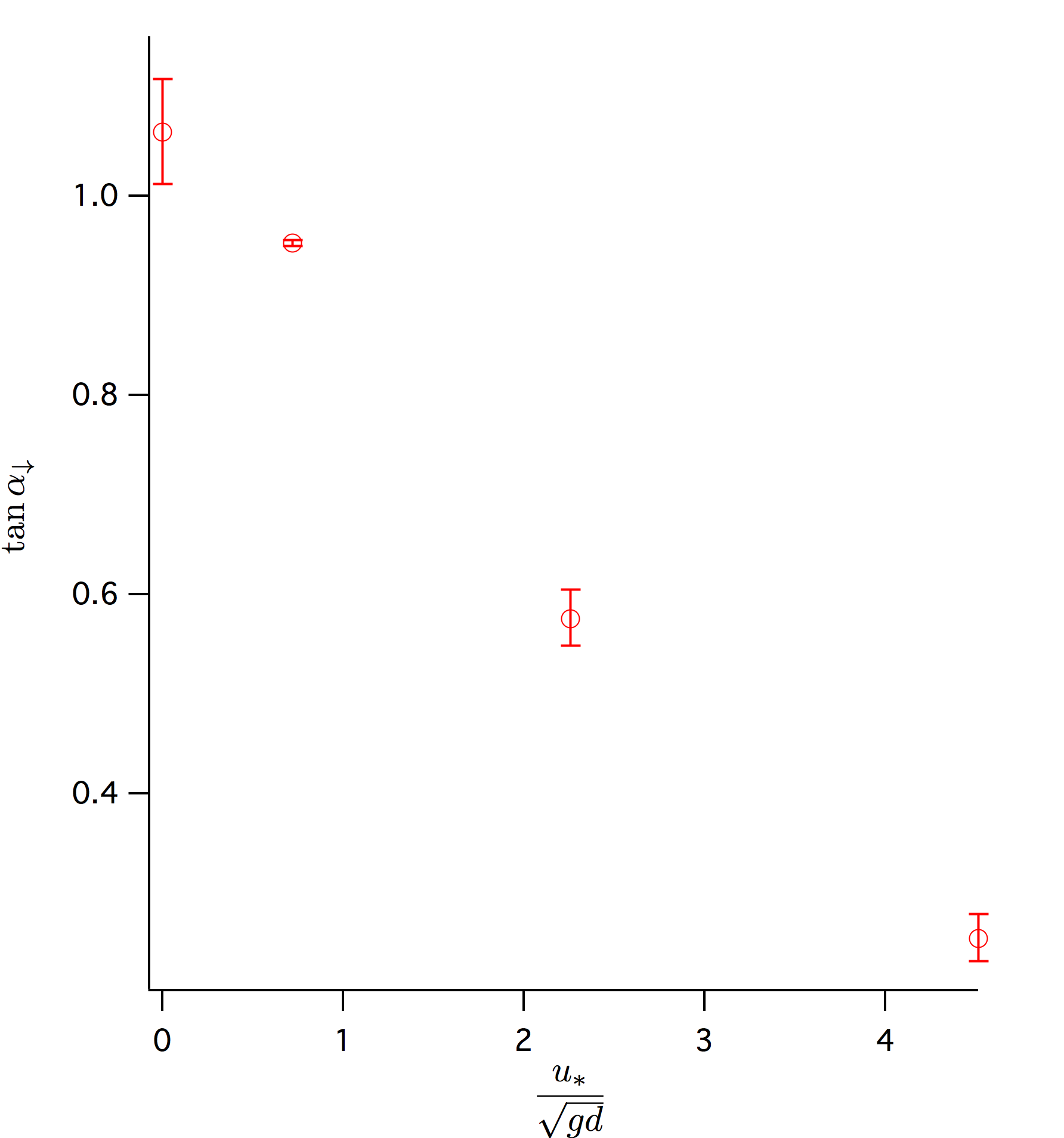}
 \caption{\footnotesize Valeur moyenne de $\tan\alpha_\downarrow$ pour différentes vitesses de cisaillement. Pour indication, on a ajouter la valeur de l'écart-type de chaque distribution centré sur la valeur moyenne. L'écart-type est à chaque fois de quelques pour cent uniquement.}
 \label{fig_dist_tanalpha}
\end{figure}

\begin{figure}[h]
 \centering
 \includegraphics[width = 10cm]{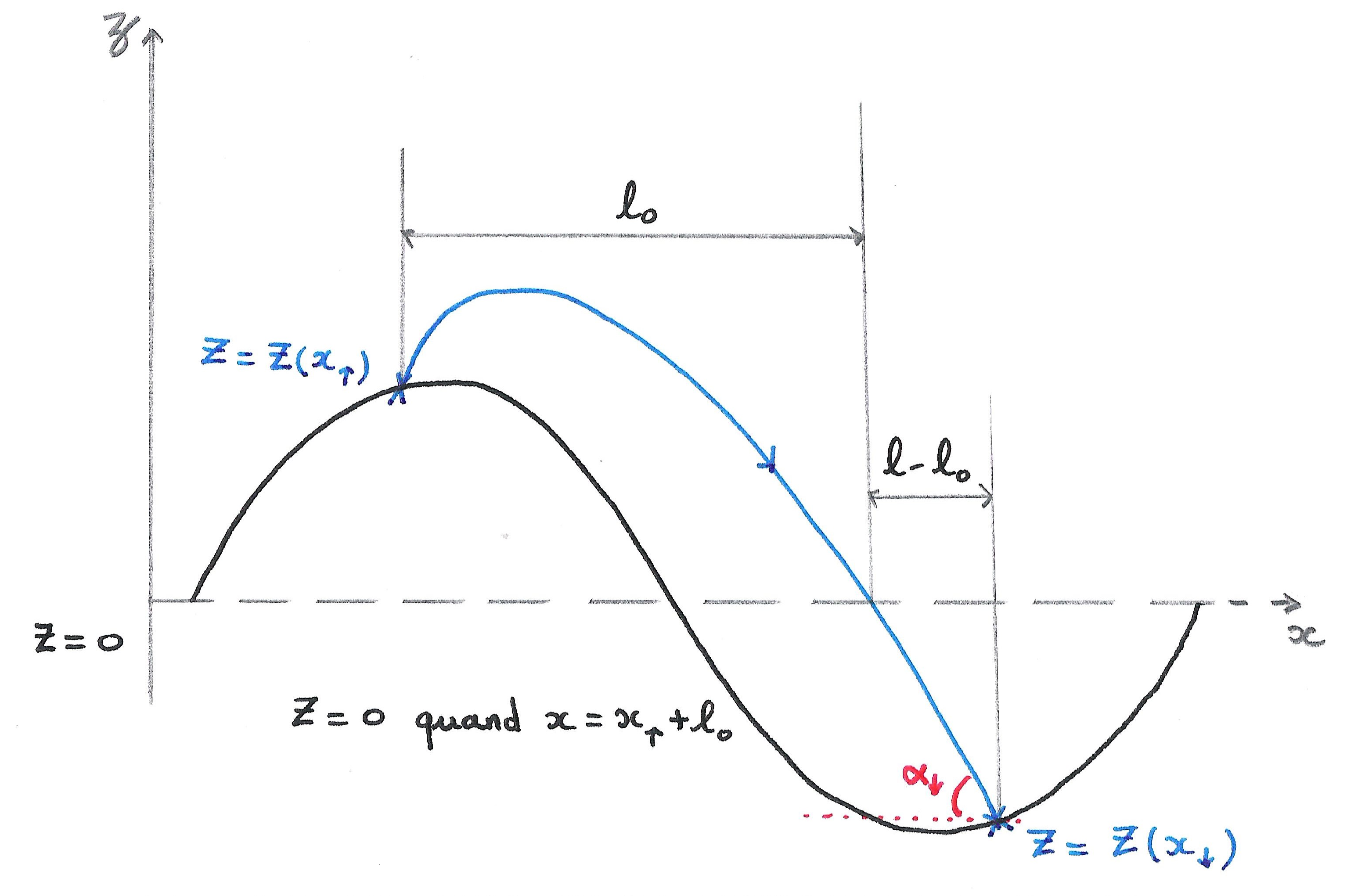}
 \caption{\footnotesize Schéma explicatif du calcul de la trajectoire d'un grain proche de l'impact, dans le cas d'un lit sinusoïdal.}
 \label{schema2}
\end{figure}

On note $\alpha_{\downarrow}$ l'angle d'arrivé du grain sur le lit (compté positivement et par rapport à l'horizontale) ; on considère cet angle indépendant de $\ell_{0}$ et constant pour un vent donné. Ceci est justifié par le fait que la distribution des $\alpha_{\downarrow}$ - pour un vent donné - est piquée autour d'une valeur moyenne (Fig. \ref{fig_dist_tanalpha}). Sur un lit plat, l'équation de la trajectoire proche de l'impact est $z = \tan \left( \alpha_{\downarrow} \right) \left( x_{\uparrow} + \ell_{0} - x \right)$. Sur un lit sinusoïdal, un grain qui part de l'altitude $Z \left( x_{\uparrow} \right)$, avec la même vitesse initiale, entre en collision avec le sol en $x_{\downarrow} = x_{\uparrow} + \ell$, où $\ell$ est la longueur de saut modifiée. En négligeant la modulation du vent induite par le relief, la trajectoire du grain est calculée de la même manière que sur lit plat et, à l'ordre linéaire en la perturbation, elle vérifie, proche de l'impact, $Z \left( x_{\downarrow} \right) = Z \left( x_{\uparrow} \right) + \tan \left( \alpha_{\downarrow} \right) \left( x_{\uparrow} + \ell_{0} - x_{\downarrow} \right)$ (Fig. \ref{schema2}). On peut réécrire cette dernière condition comme : $\ell - \ell_{0} = \frac{1}{\tan\alpha_{\downarrow}} \left[ Z \left( x_{\uparrow} \right) - Z \left( x_{\downarrow} \right) \right]$. Par identification avec l'équation (\ref{eq_variation_saut}), on en déduit l'expression de la contribution géométrique de $\mathcal{S}$ :
\begin{equation}
 \label{mathcalS-geometry}
 \mathcal{S}_{g} = \frac{1}{\tan\alpha_{\downarrow}} \left( e^{-i k \ell} - 1 \right) \, .
\end{equation}
\subsection{Effet de la modulation du vent}
\label{sec_modulation_vent}
On considère désormais l'influence de la modulation du profil de vitesse de l'écoulement. La contribution correspondante $\mathcal{S}_{w}$ intervient, numériquement, dans le calcul des trajectoires du grain à travers la force de frottement $\vec{f}$ (\ref{eq_frottements}). On choisit de décrire le champ de vitesse par le profil (\ref{eq_profil_sinusoidal}) calculé précédemment. En première approximation, les deux contribution $\mathcal{S}_{g}$ et $\mathcal{S}_{w}$ ont la même allure et leurs modules sont du même ordre de grandeur \cite{bib_model}.
\subsection{Modulation du taux de déposition}
On établie une formulation de $\psi_{1}$ à partir des différents termes du premier ordre de l'équation (\ref{eq_conservation}). Une première contribution provient de la distribution $\varphi_{1}$ que l'on évalue en $x_{\uparrow} = x_{\downarrow} - \ell$. Un second terme vient du jacobien induit par le changement de variables $(\ell_0, x_\uparrow) \leftrightarrow (\ell, x_\downarrow)$, qui s'écrit, d'après (\ref{eq_variation_saut}) :
\begin{equation}
 \mathrm{Jac} = 1 - k \zeta e^{i k x_{\downarrow}} \left( \mathcal{S}^{'} + i \mathcal{S} \right) \, ,
\end{equation}
où $\mathcal{S}^{'} = \partial \mathcal{S} / \partial \left( k \ell \right)$. Un dernier terme provient du développement $\mathcal{P}(\ell) = \mathcal{P}(\ell_{0}) + \left( \ell - \ell_{0} \right) \frac{\partial \mathcal{P}}{\partial \ell}$. Finalement :
\begin{equation}
 \label{eq_mathcalS-geometry}
 \psi_{1} \left( \ell, x_{\downarrow} \right) = P(\ell) \varphi_1 (x_\downarrow - \ell) - P(\ell) \varphi_0 \left( \mathcal{S}' + i \mathcal{S} \right ) k\zeta e^{i k x_\downarrow} - \varphi_0 \frac{ P'(\ell)}{k} \mathcal{S} k\zeta e^{i k x_\downarrow} \, .
\end{equation}
Ainsi, en utilisant les définitions (\ref{eq_varphi1}) et (\ref{eq_psi1}) de $\varphi_{1}$ et $\psi_{1}$, on détermine une relation pour $\mathcal{A}^{\psi}$ :
\begin{equation}
 \label{eq_mathcalA-general}
 \mathcal{A}^{\psi} = \mathcal{A}^{\varphi} e^{-i k \ell} - \left ( \mathcal{S}' + i \mathcal{S} \right ) - \frac{P'(\ell)}{k P(\ell)} \, \mathcal{S} \, ,
\end{equation}
avec
\begin{equation}
 \label{eq_Fuparrow-general}
 \mathcal{A}^{\varphi} =  - \frac{\int P(\ell) \left ( \mathcal{S}' + i \mathcal{S} \right ) d\ell +  \int \frac{P'(\ell)}{k} \mathcal{S} d\ell}{1 - \int P(\ell) e^{- i k \ell} d\ell} \, .
\end{equation}
L'expression analytique de $\mathcal{A}^{\psi}$, qui correspond à la contribution géométrique de (\ref{eq_mathcalS-geometry}), peut être calculée. En utilisant la condition $\int P(\ell) d\ell = 1$ et en tenant compte du fait que la distribution des longueurs de saut est bien décrite par $P(\ell) \sim 1 / \ell$, régularisée par $\ell_{min}$ et $\ell_{max}$ en $0$ et $+\infty$, l'équation (\ref{eq_Fuparrow-general}) se ramène à $\mathcal{A}^{\varphi} \sim  i/\tan\alpha_\downarrow$. On trouve finalement la relation :
\begin{equation}
 \label{eq_mathcalA-geometry}
 \tan \left( \alpha_{\downarrow} \right) \mathcal{A}_{g}^{\psi} \sim i (1+e^{-i k \ell}) + \frac{ e^{-i k \ell} - 1}{k \ell} \, ,
\end{equation}
que l'on peut écrire de manière équivalente comme :
\begin{equation}
 \label{eq_mathcalA-decomposition}
 \left\lbrace
 \begin{array}{lll}
  \Re \left[ \mathcal{A}_{g}^{\psi} \right] & = & \frac{1}{\tan\alpha_{\downarrow}} \left( \sin \left( k \ell \right) + \frac{\cos \left( k \ell \right) - 1}{k \ell} \right) \\
  \qquad \\
  \Im \left[ \mathcal{A}_{g}^{\psi} \right] & = & \frac{1}{\tan\alpha_{\downarrow}} \left( 1 + \cos \left( k \ell \right) - \frac{\sin \left( k \ell \right)}{k \ell} \right)
  \end{array}\right.
\end{equation}
\chapter{Résultats}
On cherche à déterminer les relations de dispersion $\sigma (k)$ et $\omega (k)$. Comme on l'a vu (\ref{eq_evolution}), la quantité importante à mesurer pour étudier l'évolution temporelle du lit est $\psi$. Cependant, le taux de déposition $\varphi$, qui est indépendant de $\ell$, est plus facile à manipuler. On commencera donc par considérer cette grandeur pour vérifier la consistance de nos calculs et pour déterminer la gamme de $k \zeta$ pour laquelle le régime est linéaire. 
\section{Détermination du régime linéaire}
\begin{figure}[h]
 \centering
 \includegraphics[width = 14cm]{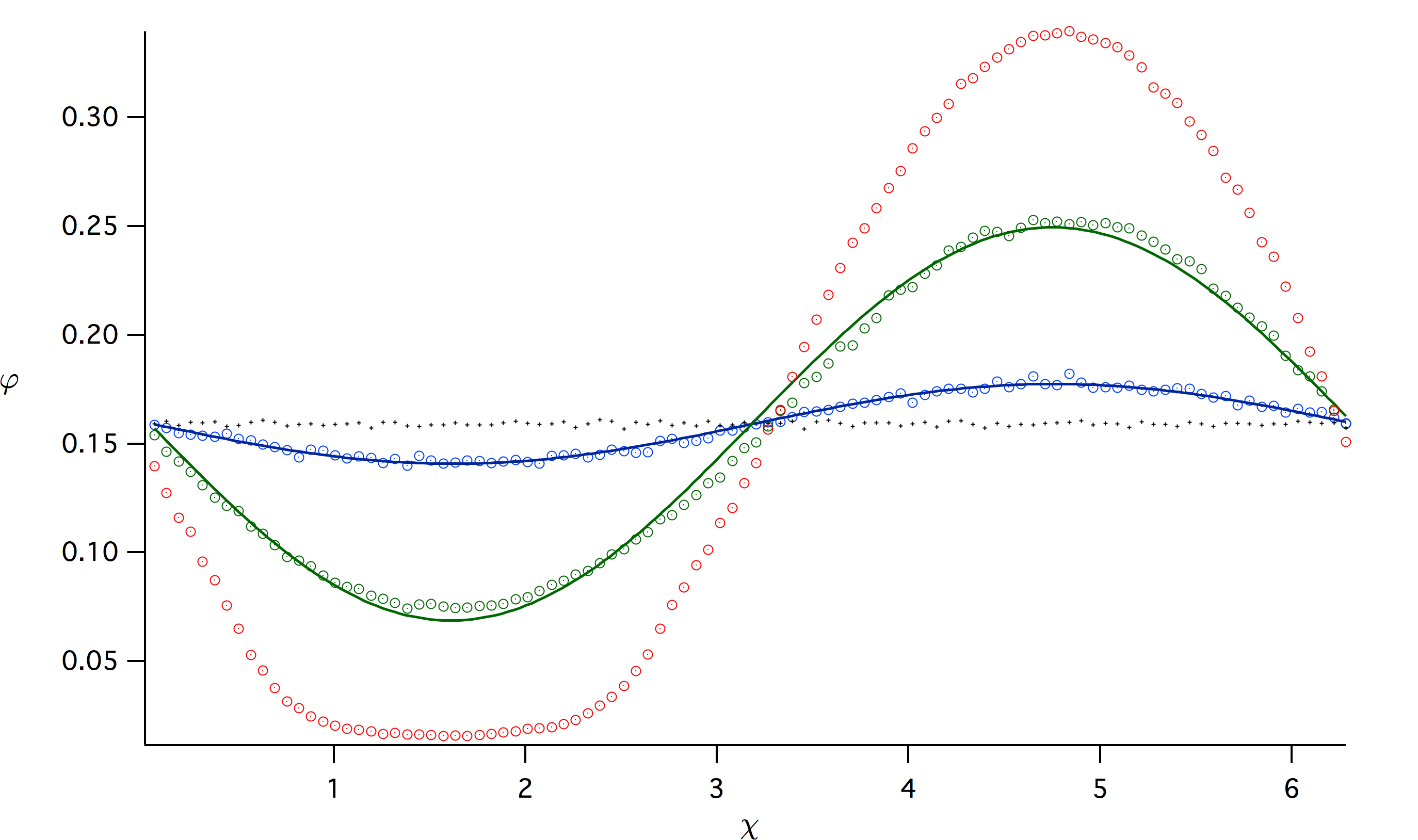}
 \caption{\footnotesize Distinction des régimes linéaire et non-linéaire : taux de déposition en fonction de la phase du lit pour différentes valeurs de $k \zeta$. En rouge $k \zeta = 0.6$ le régime est non linéaire, tandis que, en vert $k \zeta = 0.3$ et en bleu $k \zeta = 0.06$, le régime est linéaire. En noir est représenté le taux de déposition calculé sur lit plat.}
 \label{fig_regime1}
\end{figure}
\begin{figure}[h]
 \centering
 \includegraphics[width = 14cm]{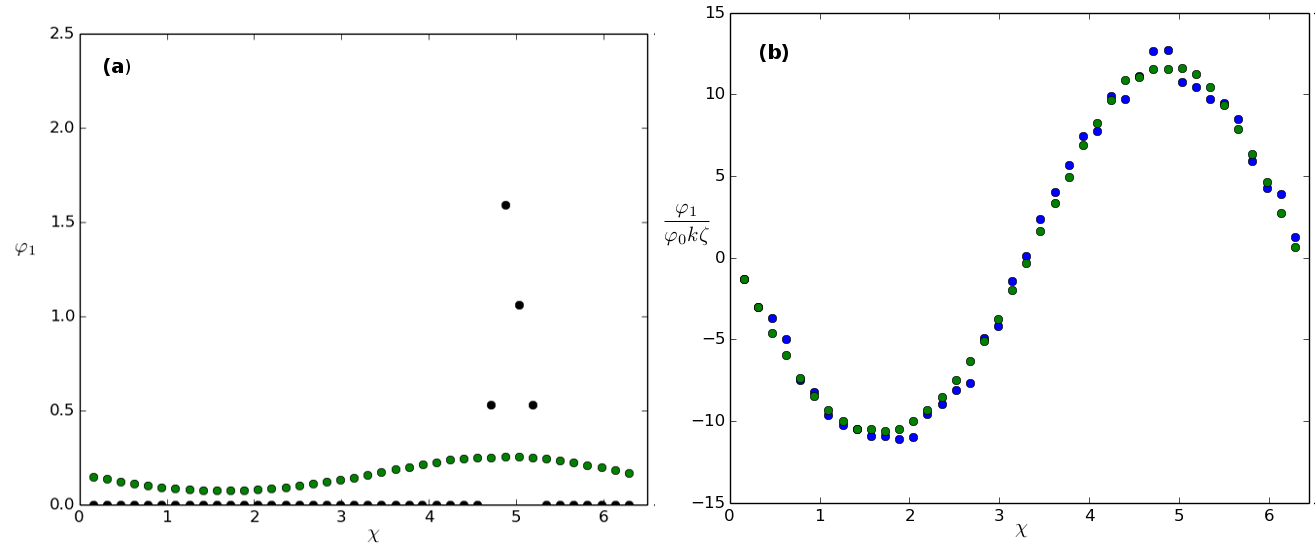}
 \caption{\footnotesize (a) Taux de déposition pour $k \zeta = 0.3$ (en vert) et $k \zeta = 1.9$ (en noir). (b) Modulation du taux de déposition divisé par $\varphi_{0} k \zeta$ pour différentes valeurs de $k \zeta$ : en vert $k \zeta = 0.3$ et en bleu $k \zeta = 0.06$.}
 \label{fig_regime2}
\end{figure}

En traçant $\varphi \left( \chi \right)$ pour différentes valeurs de $k \zeta$, on constate l'existence de deux régimes (Fig. \ref{fig_regime1}) : en dessous d'un certain facteur de forme seuil, le taux de déposition sur un lit sinusoïdal est également sinusoïdal et est proportionnel à $k \zeta$ ; on parle de régime linéaire. Au dessus du facteur de forme seuil, on parle de régime non linéaire : la dépendance en $k \zeta$ est plus complexe et $\varphi$ ne peut plus être ajusté par une fonction sinusoïdale.
Dans le cas du régime linéaire, le taux de déposition n'est pas centré sur zéro mais sur $\frac{1}{2 \pi}$. Cette valeur correspond au taux de déposition calculé sur lit plat. Ce résultat justifie la décomposition $\varphi = \varphi_{0} + \varphi_{1}$ que l'on a adoptée, où $\varphi_{1}$ est une fonction sinusoïdale de $\chi$. De plus, en déjaugeant $\varphi_{0}$ à $\varphi$ et en divisant le tout par $\varphi_{0} k \zeta$, on met en évidence la dépendance linéaire en $k \zeta$ de la modulation du taux de déposition (Fig. \ref{fig_regime2}.b).
Il existe un déphasage entre $\varphi$ et le profil du lit. Cela confirme que les grains tombent en majorité juste avant le sommet de la crête (en $\chi = \frac{3 \pi}{2}$) et en minorité juste après (en $\chi = \frac{\pi}{2}$). Pour les grandes valeurs de $k \zeta$ (régime non linéaire), les grains tendent à ne tomber plus que juste avant la crête. Dans ce cas, $\varphi$ n'est plus sinusoïdal mais est piqué autour de $\frac{3 \pi}{2}$ (Fig. \ref{fig_regime2}.a). 
On a également tracé, en trait plein sur la figure (Fig. \ref{fig_regime1}), la fonction $A^\varphi \cos(\chi +\phi^\varphi)$, où $A^\varphi$ et $\phi^\varphi$ sont les coefficients calculés à partir des projections $C^\varphi$ et $S^\varphi$. On constate que les courbes ainsi tracées s'ajustent convenablement (il y a un écart d'environ $2 \%$ pour la phase et un écart d'environ $0.02 \%$ pour l'amplitude) aux taux de déposition calculés.
\section{Module et phase de la perturbation}
\begin{figure}[h]
 \centering
 \includegraphics[width = 16cm]{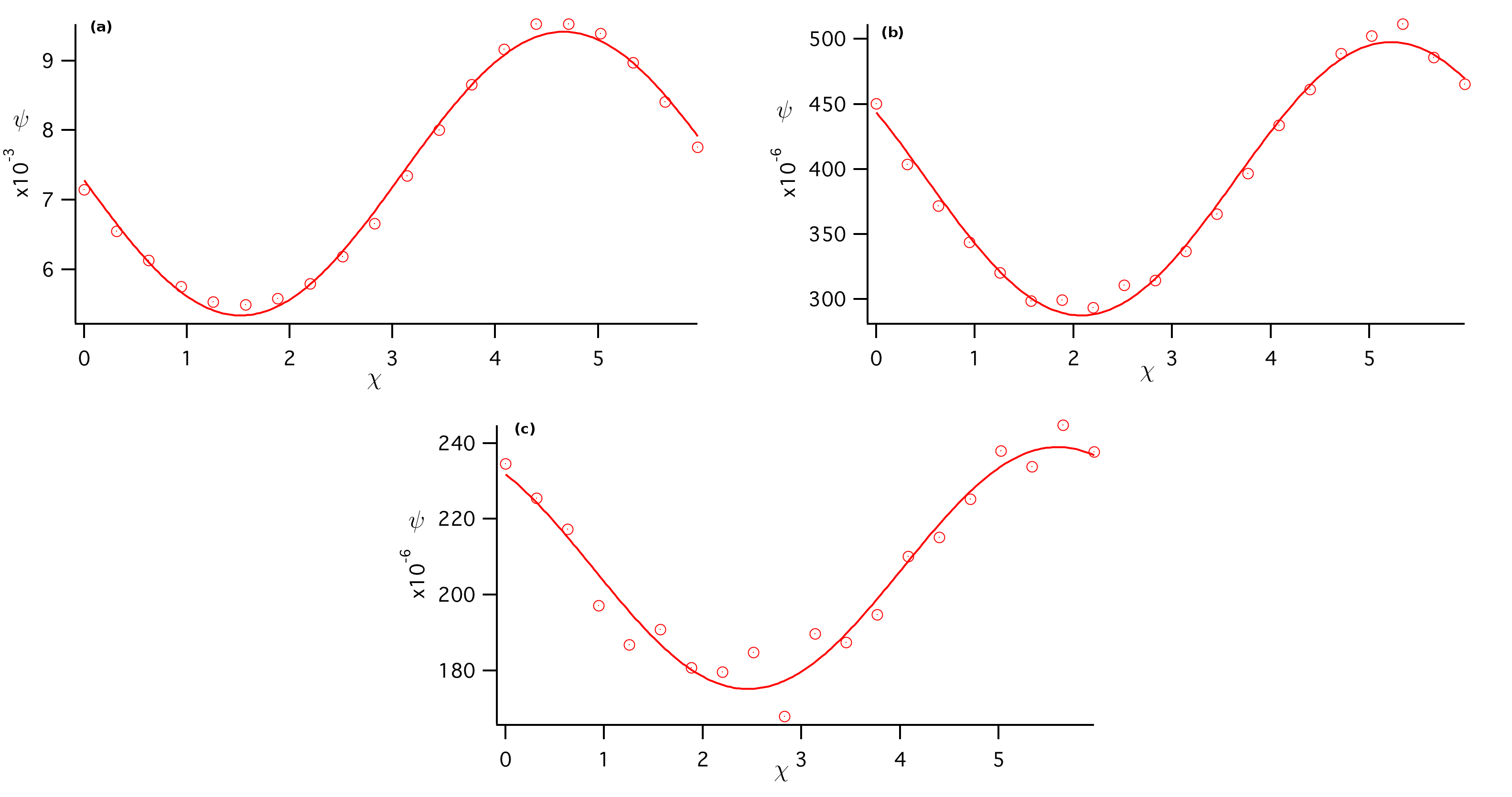}
 \caption{\footnotesize Taux de déposition $\psi \left(\chi, \ell \right)$ pour différentes longueurs de saut. (a) $\ell \in \left[ 1d \, ; 7d \right]$. (b) $\ell \in \left[ 57d \, ; 64d \right]$. (c) $\ell \in \left[ 120d \, ; 127d \right]$.}
 \label{fig_psi}
\end{figure}

\begin{figure}[p]
 \centering
 \includegraphics[width = 10cm]{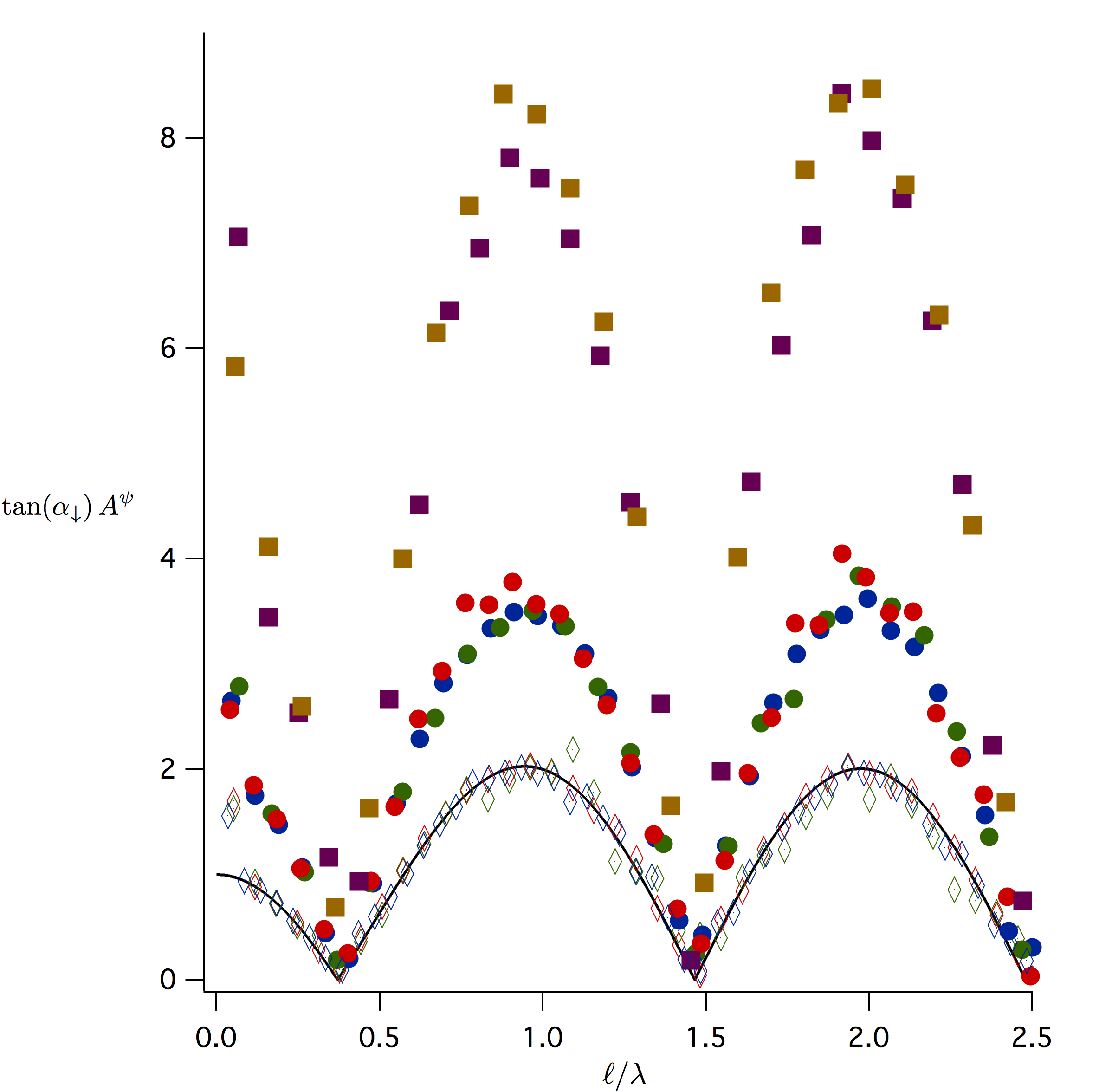}
 \caption{\footnotesize Module $A^\psi$ de la perturbation du taux de déposition limité au deux premières cloches. En trait plein noir : courbe théorique $A^\psi_g$. Les losanges correspondent aux résultats obtenus sans vent pour différentes longueurs d'onde du lit et différentes valeurs de $k \zeta$ ; les ronds correspondent aux résultats obtenus avec un vent $u_* / \sqrt{gd} \simeq 3.2$ pour différentes longueurs d'onde du lit et différentes valeurs de $k \zeta$ ; les carrés correspondent aux résultats obtenus avec un vent égal à $2u_*$ pour différentes longueurs d'onde du lit.}
 \label{fig_Apsi}
\end{figure}

On peut également vérifier que le calcul de $A^\psi$ et $\phi^\psi$ permet d'ajuster convenablement, pour une longueur de saut donnée, le taux de déposition $\psi$. On a tracé (Fig. \ref{fig_psi}) $\psi (\chi)$ pour différentes longueurs de saut. On remarque que pour les plus grands $\ell$ la qualité du signal se dégrade. Ceci est du au fait qu'il y  moins de grains qui effectuent de grands sauts que des petits. Cependant, pour une statistique suffisamment grande (basée sur $1.5*10^6$ trajectoires) on peut ajuster de manière satisfaisante $\psi$ par $A^\psi \cos \left( \chi + \phi^\psi \right)$. 
On a vu (section \ref{sec_dispersion}) que pour calculer $\sigma$ et $\omega$, il est nécessaire de connaître $A^\psi$ et $\phi^\psi$. Il a donc fallu vérifier que le calcul de ces grandeurs permet de retrouver les résultats attendus.
On a tracé (Fig. \ref{fig_Apsi}) le module de $\mathcal{A}^\psi$ en fonction de $(\ell / \lambda)$. La courbe théorique $A^\psi_g$ de $A^\psi$ - dont l'expression est calculée à partir de (\ref{eq_mathcalA-decomposition}) - se superpose parfaitement aux valeurs calculées par le modèle pour des grains non soumis à un écoulement. Par ailleurs, cette dernière courbe ayant été calculée pour un vent nul, on peut l'attribuer à la contribution géométrique de la modulation du taux de déposition ; ce qui est en accord avec l'hypothèse qui a permis d'établir sa courbe d'ajustement $A^\psi_g$. Dans ce cas, les valeurs de $A^\psi$ donnent une bonne estimation - à $\ell$ donnée - de l'intensité de la contribution géométrique.
Comme $A^\psi$ contrôle l'amplitude de $\psi_1$ et que $A^\psi$ est maximal pour les $\ell$ multiples de $\lambda$, on peut dire que la modulation du taux de déposition conditionné par la longueur de saut est maximale lorsque les grains parcourent en nombre entier de fois l'ondulation du relief. Ainsi, il existe des trajectoires raisonnantes (celles dont les longueurs de saut sont multiples de $\lambda$) pour lesquelles la réponse du taux de déposition à la perturbation du milieu granulaire est maximale.
Lorsque l'on soumet le grain à un écoulement cisaillé, la modulation du taux de déposition est d'autant plus importante que le vent est fort. Si on soustrait, à valeur de $\left( \ell / \lambda \right)$ donnée, le $A^\psi$ calculé sans vent à celui calculé avec vent, on constate que le module de $\mathcal{A}^\psi_w$, associé à la contribution du vent à la modulation de $\psi$, a la même allure que celui associé à la contribution géométrique. Ce dernier résultat est en accord avec ce qui a été trouvé par le modèle de dynamique moléculaire \cite{bib_model} et déjà souligné à la section \ref{sec_modulation_vent}. On remarque de plus que la contribution $A^\psi_w$ croit avec la vitesse $u_*$ de l'écoulement.
\begin{figure}[p]
 \centering
 \includegraphics[width = 14cm]{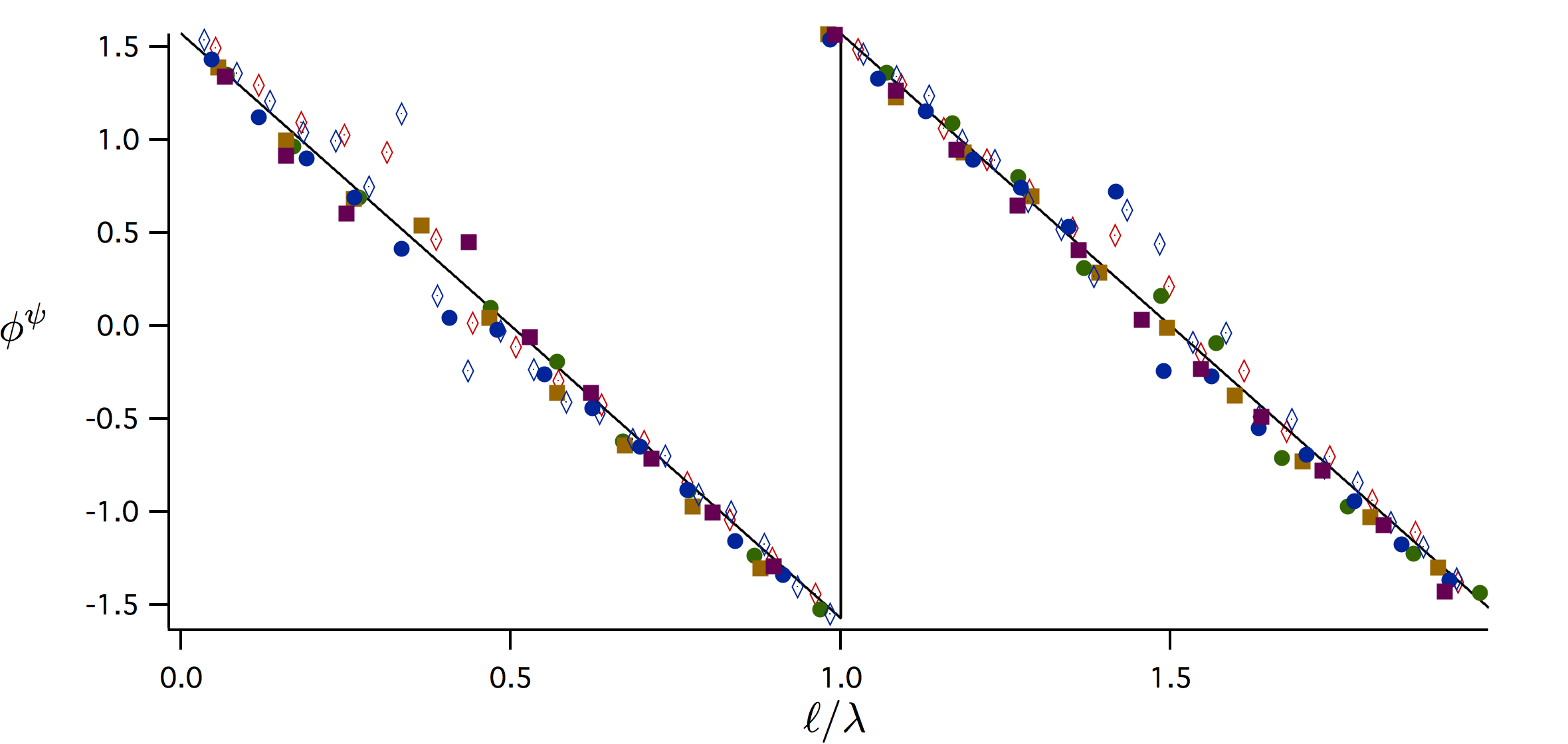}
 \caption{\footnotesize Phase $\phi^\psi$ de la modulation du taux de déposition. En trait plein noir : courbe théorique $\phi^\psi_g$. Les losanges correspondent aux résultats obtenus sans vent pour différentes longueurs d'onde du lit ; les ronds correspondent aux résultats obtenus avec un vent $u_* / \sqrt{gd} \simeq 3.2$ pour différentes longueurs d'onde du lit et différentes valeurs de $k \zeta$ ; les carrés correspondent aux résultats obtenus avec un vent égal à $2u_*$ pour différentes longueurs d'onde du lit.}
 \label{fig_phi}
\end{figure}
Le calcul de la phase de la perturbation par le modèle simplifié permet de retrouver le résultat théorique attendu (calculé à partir de (\ref{eq_mathcalA-decomposition})). On constate (Fig. \ref{fig_phi}) que, contrairement au module de la perturbation, $\phi^\psi$ ne dépend pas de la force de l'écoulement. De plus, lorsque la réponse à la perturbation est maximale - ie $\ell$ multiple de $\lambda$ - le déphasage avec le lit vaut $\frac{\pi}{2}$. Comme le profil du lit est proportionnel à $\cos \left( \chi \right)$, ce déphasage traduit que le maximum de grains arrivent en $2 \pi - \frac{\pi}{2} = \frac{3 \pi}{2}$ ; soit juste avant le sommet de la crête.
\section{Taux de croissance}
On a, dans un premier temps, essayé de déterminer le taux de croissance en fixant la longueur de coupure $l_c$ et en jouant sur $\varepsilon_g$ ; $\varepsilon_d$ est alors fixé par la condition sur lit plat (\ref{lit_plat}). Pour choisir $l_c$ de manière cohérente, on a étudié la distribution des angles d’impact $\alpha_\downarrow$, avec l’horizontale, du grain incident. Sur lit plat, sans vent et en négligeant les frottements avec le fluide, un grain effectue des trajectoires paraboliques. La prise en compte des frottements et du vent contribue à déformer les trajectoires en les allongeant dans la direction de l’écoulement. Ainsi, un grain soumis à un écoulement cisaillé arrivera sur le lit avec un angle $\alpha_\downarrow$ plus petit que celui d’un grain identique - dont la trajectoire est initiée avec les mêmes paramètres - non soumis au vent. La conclusion est la même pour un lit de profil sinusoïdal.
On a défini les saltons comme les grains qui disposent de suffisamment d’énergie pour rebondir et déloger d’autres particules lors de l’impact. On a défini les reptons comme les grains qui sont éjectés, lors le l’impact d’un salton avec le lit, et qui ne sont pas eux-mêmes suffisamment énergétiques pour rebondir suite à leur saut. Dans notre modèle, on peut donc associer les grains qui effectuent des petits sauts aux reptons et ceux qui en font des grands aux saltons. Pour fixer la limite entre ces deux régimes, on qualifiera de grand saut la longueur de saut des grains dont la trajectoire a le temps d’être modifiée par l’action de l’écoulement (c’est-à-dire dont l’angle d’impact est aplati par rapport au cas sans vent) et de petit saut la longueur de saut des grains dont la trajectoire n’est pas influencée par le mouvement du fluide environnant. Ainsi, $l_c$ correspond à la longueur caractéristique pour laquelle les grains commencent à ressentir l’action du vent. On a trouvé $l_c \simeq 40d$.
Cependant, cette approche ne permet pas de trouver un résultat satisfaisant. En effet, la détermination des deux autres paramètres associés à $\varepsilon$, tels que $\sum w_n$ converge, est difficile et $\varepsilon$ est elle-même trop sensible à la variation de ces grandeurs. Dans ces conditions, on ne retrouve pas un $\sigma$ conforme aux observations. \medskip
On adopte une autre approche qui consiste à fixer $\varepsilon_g$ et à jouer avec la longueur de coupure. Il faut alors choisir $l_c$ telle que $\sigma$ soit positif lorsqu'il y a déposition. D'après (\ref{eq_sigma1}), le taux de croissance s'écrit, dans le cas d'une fonction $\varepsilon (\ell)$ de la forme (\ref{eq_creneau}) : 
\begin{eqnarray}
 \sigma (k) & = & \frac{k \varphi_0}{\tan \alpha_\downarrow} \left[ \int_0^{\ell_c} \left( \frac{\sin \left( k \ell \right)}{\ell} + \frac{\cos \left( k \ell \right) - 1}{k \ell^2} \right) \varepsilon_g d\ell \right. \nonumber \\
 \qquad
 &  & + \left. \int_{\ell_c}^{+\infty} \left( \frac{\sin \left( k \ell \right)}{\ell} + \frac{\cos \left( k \ell \right) - 1}{k \ell^2} \right) \varepsilon_d d\ell \right] \, . \nonumber
\end{eqnarray}
Soit, en effectuant le changement de variable $\varsigma = k \ell$ :
\begin{eqnarray}
 \label{eq_sigma3}
 \sigma \left( k \right)& = & \frac{k \varphi_0}{\tan \alpha_\downarrow} \left[ \int_0^{k \ell_c} \left( \frac{\sin \left( \varsigma \right)}{\varsigma} + k \frac{\cos \left( \varsigma \right) - 1}{\varsigma^2} \right) \varepsilon_g d\varsigma \right. \nonumber \\
 \qquad
 &  & + \left. \int_{k \ell_c}^{+\infty} \left( \frac{\sin \left( \varsigma \right)}{\varsigma} + k \frac{\cos \left( \varsigma \right) - 1}{\varsigma^2} \right) \varepsilon_d d\varsigma \right] \, .
\end{eqnarray}
Le modèle de dynamique moléculaire nous a appris que la croissance des rides éoliennes est dominée par l'action des reptons. Autrement dit, l'intégrale qui donne le taux de croissance est dominée par les petits $\ell$ relativement à la séparation des régimes de transport par saltation et de transport par reptation. Pour que cette condition soit respectée, il faut que $k \ell_c$ soit tel que l'intégrale de $\sigma \left( k \right)$ soit contrôlée par la partie positive de la première oscillation du sinus cardinal $\frac{\sin \left( \varsigma \right)}{\varsigma}$ (présent dans la première intégrale de (\ref{eq_sigma3})). Pour ce faire, il est nécessaire que $k \ell_c$ soit constant d'ordre 1. On identifie alors les petites longueurs de saut aux sauts tels que $k \ell \lesssim 1$.
Cependant, cette manière de procéder est artificielle car on force nous même une des bornes des intégrales avec $k \ell_c$ pour obtenir le résultat attendu pour $\sigma$. Or, la coupure entre les régimes de saltation et de reptation n'a, a priori, aucune raison de dépendre de $k$. En effet, la capacité de remplacement $\left( 1 - \varepsilon \right)$ correspond à la fraction de grains réémis après chaque impact d'une particule incidente et ne dépend a priori que de l'énergie du grain et éventuellement de la géométrie du lit au point d'impact.
Par ailleurs, $\varepsilon$ doit être une fonction sans dimension. On l'a défini jusqu'à présent comme dépendante d'une longueur : $\ell_c$. Pour que $\varepsilon$ soit effectivement sans dimension, il faut qu'elle dépende d'un nombre sans dimension, c'est-à-dire d'un rapport de longueurs. Plusieurs solutions sont imaginables. On peut par exemple supposer une dépendance en $\ell / d$, ou une dépendance en $\ell_i / \ell_e$ avec $\ell_i$ la longueur du saut que le grain incident a effectué et $\ell_e$ la longueur de saut que le grain émis a effectué. Ces hypothèses, si elles sont vérifiées, ne sont à l'heure actuelle pas comprises.
\chapter*{Conclusion}
\addcontentsline{toc}{chapter}{Conclusion}
Le seul fait que l'on ait pu retrouver le bon comportement de $A^\psi$ et $\phi^\psi$ confirme la légitimité et l'utilité d'un modèle simplifié du transport sédimentaire. En effet, le modèle que l'on a développé permet de retrouver les bons comportements de transport sédimentaire et de déposition, uniquement à partir des équations simplifiées de la dynamique et des lois de conservation de la matière. En outre, la possibilité d'utiliser une description statistique des processus physiques pour décrire les phénomènes est très satisfaisante puisqu'elle permet de retrouver les bonnes lois d'échelle tout en minimisant les temps de calcul. La durée plus courte des simulations offre la possibilité de faire varier plus facilement les paramètres de contrôle (rapport des masses volumiques, vitesse de l'écoulement, etc.) pour explorer qu'elles sont les grandeurs qui caractérisent le développement des rides éoliennes. On a ainsi pu constater que la modulation du taux de déposition fluctue davantage, d'une longueur de saut à une autre, lorsque l'écoulement est plus puissant, mais que le déphasage avec le profil du lit n'en dépend pas.
On a uniquement considéré, dans notre étude, des lits sinusoïdaux de différentes longueurs d'onde, soumis à un écoulement cisaillé. En effet, en identifiant les ondulations aux rides éoliennes formées sur un lit de sable et en étudiant les taux de déposition $\psi \left( \chi, \ell \right)$ associés, on peut établir les relations de dispersion qui caractérisent l'évolution du milieu granulaire. À chaque longueur d'onde $\lambda$ correspond une certaine étape de l'évolution du lit et le calcul de $\sigma (k)$ permet de déterminer quelles sont les $\lambda$ privilégiées lors du développement des rides éoliennes. Dans l'optique de minimiser la complexité de notre modèle, on a souhaité parvenir à la détermination du taux de croissance en introduisant une fonction $\varepsilon$ représentative de la fraction moyenne de grains réémis lors de l'impact d'un salton avec le lit. Comme on l'a vu, le choix des paramètres qui définissent $\varepsilon$ n'est pas évident. Notamment, l'introduction de la notion de longueur de coupure - qui caractérise le passage du transport par saltation au transport par reptation - n'est pas intuitive. De plus, la grande sensibilité de $\varepsilon$ à la modification de ses paramètres rend le calcul de $\sigma$ difficile.
Afin d'achever l'ajustement du modèle simplifié au modèle de dynamique moléculaire, on doit donc trouver les paramètres $\varepsilon_d$, $\varepsilon_g$ et $l_c$ (définissant $\varepsilon$) qui permettent d'obtenir le bon comportement de $\sigma (k)$. Il faut ensuite vérifier que ce choix, qui permet de calculer les bonnes valeurs de $\sigma$, est reproductible lorsque l'on modifie les paramètres de contrôle et qu'il permet de trouver une vitesse de propagation $\omega / k$ correcte. Dans le cas où on ne pourrait pas reproduire le taux de croissance calculé par le modèle de dynamique moléculaire, il faudrait explorer d'autres fonctions $\varepsilon$ plus complexes que la fonction créneau dans leur forme et leurs dépendances. Puisque l'on désire travailler avec un modèle simplifié, ces fonctions devront être intégrables de manière analytique pour que l'on puisse les incorporer facilement au modèle. Si toutefois on ne parvenait toujours pas à reproduire les bonnes relations de dispersion, il faudrait conclure que la représentation des processus - qui régissent l'évolution des rides éoliennes - par une fonction $\varepsilon$ n'est pas compatible avec les observations.
Pour aller plus loin, une fois que le modèle permettra de retrouver les relations de dispersion, on pourra considérer l'\textit{évolution dynamique} du milieu granulaire - par opposition à l'évolution image par image, lorsque l'on considérait des profils sinusoïdaux de différentes longueurs d'onde - en prenant compte de la déformation du lit sous l'impact des saltons et de la déposition des reptons. Toujours en considérant des échelles de temps différentes pour le transport et l'évolution du relief, on pourra étudier statistiquement la formation et l'évolution des rides éoliennes sur un milieu granulaire. On pourra, par exemple, associer aux saltons une \textit{fonction cratère} dépendante de l'énergie du grain incident et représentative du nombre de grains éjectés. L'étude de tels systèmes, pour différents paramètres de contrôle, devrait permettre d'isoler les grandeurs qui dominent les processus d'évolution des rides et ainsi mieux comprendre ce phénomène. 
\newpage
\pagestyle{empty}
\null
\newpage
\bibliographystyle{unsrt}
\bibliography{compte_rendu}
\end{document}